\documentclass[11pt]{article}
\usepackage[utf8x]{inputenc}
\usepackage[T1]{fontenc}
\usepackage[margin=1.in,marginparwidth=1.75cm]{geometry}
\usepackage{amsmath, graphicx, amsthm, amssymb, bbm, enumerate,complexity}
\usepackage[ruled]{algorithm2e}
\usepackage[pagebackref,colorlinks,citecolor=blue,urlcolor=magenta,linkcolor=magenta,,bookmarks=true]{hyperref}

\usepackage{stmaryrd}
\usepackage{xcolor}
\usepackage[nameinlink,capitalize]{cleveref}
\usepackage{mathtools}
\usepackage{amsmath,amssymb,amsfonts,amsthm,epsfig}

\usepackage{bm,xspace}

\usepackage{cancel}
\usepackage{fullpage}
\usepackage{framed}
\usepackage{verbatim}
\usepackage{array}
\usepackage{multirow}
\usepackage{afterpage}
\usepackage{mathrsfs}  

\title{New Pseudorandom Generators and Correlation Bounds Using Extractors}

\author{Vinayak M. Kumar
\thanks{Department of Computer Science, University of Texas at Austin. Email: {\tt vmkumar@utexas.edu}. Supported by NSF Grant CCF-2008076, CCF-2312573, and a Simons Investigator Award (\#409864, David Zuckerman).}}
\date{}

\newcommand{\corr}{\func{corr}}

\newcommand{\lsb}{\func{lsb}}

\newcommand{\junta}{\func{JUNTA}}

\newcommand{\OR}{\func{OR}}
\newcommand{\Ext}{\func{Ext}}

\newcommand{\extffm}{\func{ExtFFM}}
\newcommand{\ffm}{\func{FFM}}
\newcommand{\tc}{\func{TC}}
\newcommand{\AND}{\func{AND}}
\newcommand{\NOT}{\func{NOT}}

\newcommand{\sym}{\func{SYM}}

\newcommand{\thr}{\func{THR}}

\newcommand{\gip}{\func{GIP}}
\newcommand{\rw}{\func{RW}}
\newcommand{\pargate}{\func{PAR}}

\newcommand{\g}{\func{G}}

\newcommand{\ac}{\func{AC}^0}

\newcommand{\bits}{\{0,1\}}
\newcommand{\calF}{{\cal F}}

\newcommand{\calC}{{\cal C}}

\newcommand{\calE}{{\cal E}}
\newcommand{\calR}{{\cal R}}

\renewcommand{\E}{\mathbb{E}}

\newcommand{\F}{\mathbb{F}}

\newcommand{\eps}{\varepsilon}

\newcommand{\conj}{\overline}

\newtheorem{thm}{Theorem}[section]
\newtheorem*{remark}{Remark}

\newtheorem{cor}[thm]{Corollary}
\newtheorem{lem}[thm]{Lemma}
\newtheorem{defn}[thm]{Definition}
\newtheorem{claim}[thm]{Claim}

\makeatletter
\newcommand{\ostar}{\mathbin{\mathpalette\make@circled\star}}
\newcommand{\make@circled}[2]{%
  \ooalign{$\m@th#1\smallbigcirc{#1}$\cr\hidewidth$\m@th#1#2$\hidewidth\cr}%
}
\newcommand{\smallbigcirc}[1]{%
  \vcenter{\hbox{\scalebox{0.77778}{$\m@th#1\bigcirc$}}}%
}
\makeatother
\begin{document}

\maketitle

\begin{abstract}
    We establish new correlation bounds and pseudorandom generators for a collection of computation models. These models are all natural generalizations of structured low-degree $\F_2$-polynomials that we did not have correlation bounds for before. In particular: 
    \begin{itemize}
        
        \item We construct a PRG for width-2 $\poly(n)$-length branching programs which read $d$ bits at a time with seed length $2^{O(\sqrt{\log n})}\cdot d^2\log^2(1/\eps)$. This comes quadratically close to optimal dependence in $d$ and $\log(1/\eps)$. Improving the dependence on $n$ would imply nontrivial PRGs for $\log n$-degree $\F_2$-polynomials. The previous PRG by Bogdanov, Dvir, Verbin, and Yehudayoff had an exponentially worse dependence on $d$ with seed length of $O(d\log n + d2^d\log(1/\eps))$.
        
        \item We provide the first nontrivial (and nearly optimal) correlation bounds and PRGs against size-$n^{\Omega(\log n)}$ $\ac$ circuits with either $n^{.99}$ $\sym$ gates (computing an arbitrary symmetric function) or $n^{.49}$ $\thr$ gates (computing an arbitrary linear threshold function). This is a generalization of sparse $\F_2$-polynomials, which can be simulated by an $\ac$ circuit with one parity gate at the top. Previous work of Servedio and Tan only handled $n^{.49}$ $\sym$ gates or $n^{.24}$ $\thr$ gates, and previous work of Lovett and Srinivasan only handled polynomial-size circuits.
        
        \item We give exponentially small correlation bounds against degree-$n^{O(1)}$ $\F_2$-polynomials which are set-multilinear over \emph{some} arbitrary partition of the input into $n^{.99}$ parts (noting that at $n$ parts, we recover \emph{all} low-degree polynomials). This vastly generalizes correlation bounds against degree-$(d-1)$ polynomials which are set-multilinear over a fixed partition into $d$ blocks, which were established by Bhrushundi, Harsha, Hatami, Kopparty and Kumar.
    
    \end{itemize}

    The common technique behind all of these results is to fortify a hard function with the right type of extractor to obtain stronger correlation bounds for more general models of computation. Although this technique has been used in previous work, it relies on the model shrinking to a very small computational class under random restrictions. We view our results as a proof of concept that such fortification can be done even for classes that do not enjoy such behavior.
\end{abstract}
\break 
\tableofcontents
\break

\section{Introduction/Outline of Results}

Many central questions in complexity theory revolve around proving limitations of various computational models. For example, there are research programs which seek lower bounds against constant depth circuits, low-degree polynomials over $\F_2$, and perhaps most famously the complexity class $\func{P}$.\ 

Usually, lower bounds against a simple class of $n$-bit Boolean functions $\calC$ are established by demonstrating an explicit function $f$ such that no $g\in\calC$ can compute $f$ on every input. This is referred to as \emph{worst-case hardness}. However, we may not be satisfied with this in practice and stipulate that no $g\in\calC$ can even \emph{approximate} $f$. After all, if there exists a $g$ that agrees with $f$ on all but one point, the difference may be impossible to detect in practice. Furthermore, establishing average case hardness against $\calC$ can allow us to create PRGs against $\calC$ via the ``hardness to randomness'' framework introduced by Nisan and Wigderson \cite{nw94}, as well as show hardness results against related function classes, like the majority of functions in $\calC$. This \emph{average-case hardness} statement is exactly what the study of correlation bounds capture.  

To formally define this, let $D$ be a distribution over $\bits^n$. Define the \emph{correlation} of two Boolean functions $f,g : \bits^n\to\bits$ \emph{over} $D$ to be 
\[\corr_D(f,g) = |\E_{x\sim D }[(-1)^{f(x)+g(x)}]|.\] We will usually be concerned with $D = U_n$, the uniform distribution, and should be assumed so if no distribution $D$ is specified. Notice that this quantity is a real number in $[0,1]$. For intuition, note that if $f=g$ or $f = 1-g$, the correlation is 1, whereas if $f$ and $g$ only match on about half the inputs, the correlation becomes small. This fact allows us to observe that correlation is the right notion, as $\corr(f,g)$ being small implies that $g$ cannot predict $f$ much better than a coin flip. For a function $f$ and a function class $\calC$, we can define $\corr(f,\calC) = \max_{g\in\calC}\corr(f,g)$. Hence, the notion of $f$ being average-case hard for $\calC$ is captured by $\corr(f,\calC)$ being small.  \ 

In this paper, we are most interested in the case where $\calC$ is the class of low-degree $\F_2[x_1,\dots, x_n]$ polynomials. Establishing correlation bounds against low-degree $\F_2$ polynomials is an extremely interesting and central question in complexity theory, as it is either necessary or sufficient to understand a plethora of other problems, some of which concern communication protocols, matrix rigidity, and PRGs for circuits. See Viola's survey \cite{vio22} for a detailed exposition on this rich program. 

Unfortunately, there is a ``$\log n$-degree barrier'' for PRGs and correlation bounds against low-degree polynomials. Current PRGs and correlation bounds are asymptotically tight for constant degree polynomials, but become \emph{trivial} at degree $\log n$ \cite{vio08}. Getting nontrivial PRGs (or even correlation bounds) against $\log n$-degree polynomials has been a longstanding open problem.

Towards breaking this barrier, researchers have shown strong correlation bounds for \emph{structured} subsets of low-degree $\F_2$-polynomials (such as sparse polynomials \cite{ls11, st19}, tensors \cite{bhhkk}, small-read polynomials, and symmetric polynomials \cite{bijlsv21}) with the hope of being able to generalize them. In this work, we establish new correlation bounds and PRGs for computation models generalizing some of these polynomials, namely width-2 branching programs reading $d$ bits at a time, $\ac$ containing a small number of arbitrary symmetric or linear threshold gates, and set-multilinear polynomials.

Interestingly, all of these correlation bounds are obtained by taking a function hard for a more specific class of polynomials and then \emph{fortifying} it with a well-suited extractor. Although such a strategy is not new and has been used to establish stronger lower bounds for formulas \cite{krt13,ckksz14}, they usually rely on the fact that upon randomly fixing a subset of variables of a formula, there are extremely few possibilities for the resulting function. Our work shows that extractor fortification is a much broader technique that can strengthen lower bounds against function classes even if they do not simplify to a tiny collection of functions under a random restriction. In particular, our correlation bounds demonstrate that extractor fortification can work as long as the function class, after a random restriction, has low communication complexity or good algebraic structure. 

Inspired by this, we would like to understand when extractors will strengthen correlation bounds. Due to technical reasons, this seems nontrivial to establish. We explain this technical nuance in \Cref{sec:nuance}. 

The remainder of this section is devoted to introducing and motivating each computational model studied, surveying prior work in the topic, and stating all key results proven.

\subsection{Better Bounds and PRGs Against $\ac$ with More $\{\sym,\thr\}$ Gates}

 Our knowledge of hardness and PRG results for $\ac$ is much more developed than that of $\tc^0$. Our state of the art PRGs for $\ac$ is Lyu's construction \cite{Lyu}, which $\eps$-fools polysize $\ac$ circuits with seed length $\tilde{O}(\log^{d-1}(n)\log(n/\eps))$, whereas the current best PRG of Hatami, Hoza, Tal, and Tell which $(2^{-n^\delta})$-fools size- $O(n^{1+\delta})$ $\tc^0$ circuits have seed length $O(n^{1-\delta})$ \cite{hhtt21}. Due to this stark contrast in parameters, it is natural to gradually work upward from $\ac$ by allotting a budget of $\sym$ (calculates an arbitrary symmetric function) or $\thr$ (calculates an arbitrary linear threshold function) gates in the circuit. This approach has been explored for more than a decade \cite{v07,ls11,st19}, building upon the study of PRGs for $\{\sym,\thr\}\circ\ac$ circuits pioneered by Luby, Velickovi\'{c}, and Wigderson \cite{lvw93}. This context explains why this circuit class a compelling generalization of sparse polynomials (which can be written as a small-size parity of ands). All the mentioned works use the following function introduced by Razborov and Wigderson in 1993 \cite{rw93} (all arithmetic is over $\F_2$).

 \begin{align}\rw_{m,k,r}(x) = \sum_{i=1}^m\prod_{j=1}^k\sum_{\ell = 1}^r x_{ij\ell} \label{eqn:rw}\end{align}

Most recently, Servedio and Tan \cite{st19} use $\rw_{m,k,r}$ to uncorrelate against constant-depth size-$n^{O(\log n)}$ $\ac$ circuits whose top gate is $\{\sym,\thr\}$ (denoted $\{\sym,\thr\}\circ \ac$). Their explicit bound is \[\corr\left(\rw_{\sqrt{\frac{n}{\log n}},\log n, \sqrt{\frac{n}{\log n}}}, \; \{\sym,\thr\}\circ\ac\right)\le 2^{-\Omega(n^{.499})}.\] By the techniques used in \cite{ls11}, this can be translated into correlation bounds against $\ac$ circuits with up to $n^{.499}$ $\sym$ gates or $n^{.249}$ $\thr$ gates. As can be seen by the repeated occurrences of $n^{.499}$, the strength of the correlation bound dictates how many $\{\sym,\thr\}$ gates we can afford in our budget. \ 

We show that $\rw$ is just one of many functions from a general class of hard functions with small correlation against $\{\sym,\thr\}\circ\ac$ circuits.  For functions $f:(\bits^r)^k\to \bits$ and $g: \bits^m\to\bits^r$, denote $f\circ g^k(x_1\dots, x_k):= f(g(x_1),\dots g(x_k)).$

\begin{thm}[informal]
    Let $g$ be computable by a size $n^{O(\log n)}$ $\{\sym,\thr\}\circ \ac$ circuit. Let $f$ be average-case hard against multiparty protocols\footnote{the formal condition is any function with small ``$k$-party norm'' or ``cube norm'', but this is currently the only technique we know that establishes average case hardness against multiparty protocols.}, and let $\Ext$ be a suitable extractor. Then \[\corr(f\circ \Ext^{.01\log n }, g)\le 2^{-\Omega(n^{.999})}.\]
\end{thm}

To our knowledge, this theorem gives the first context in which generically precomposing with an extractor boosts correlation bounds whose proof does not rely on simplification under random restriction (indeed, parity does not simplify under restriction and is contained in $\{\sym,\thr\}\circ \AC^0\}$). Previously, extractors have only been used to boost correlation bounds for classes that heavily simplify under random restriction \cite{krt13, ckksz14, css16, kkl17}, whether it be a small depth decision tree or a function very close to a constant.\footnote{There have been uses of extractors as a hard function against classes that do not simplify under restriction, like DNFs of Parities \cite{cs16dnfofparities} and strongly read-once
linear branching programs \cite{OGrolbp,lz24rolbp,cl23ROLBP}. However, they directly establish a correlation bound against the extractor rather than amplify a weaker hard function by precomposing with an extractor.} Our theorem states that extractors can still boost correlation bounds, even if they were proven using communication complexity or algebraic methods (rather than only random restrictions).

Furthermore, our theorem distills the reason why $\rw$ was so effective as a hard function. Quantitatively, we can instantiate the template with a suitable extractor to obtain a new hard function with nearly-optimal correlation bounds.

Due to our strengthened correlation bounds, we can now obtain correlation bounds and PRGs against size-$n^{O(\log n)}$ $\ac$ circuits with up to $n^{.999}$ $\sym$ gates or $n^{.499}$ $\thr$ gates. Prior to this, no nontrivial correlation bound or PRG was known to handle such large size \emph{and} number of $\{\sym,\thr\}$ gates (\cite{ls11} could handle the same number of $\{\sym,\thr\}$ gates but only for $n^{O(\log\log n)}$-size circuits, and \cite{st19} could handle the same size circuits, but only $n^{.499}$ $\sym$ or $n^{.249}$ $\thr$ gates).

Even for $\{\sym,\thr\}\circ \ac$ circuits that have only one $\{\sym,\thr\}$ gate, our correlation bounds yield improved PRGs whose seed length is $2^{O(\sqrt{\log S})} + (\log(1/\eps))^{2.01}$, which has a better dependence on $\eps$ than in previous work (see Table 1). In fact, since the best correlation bound that one can hope for is $2^{-\Omega(n)}$, this dependence is almost optimal under the Nisan-Wigderson framework, and an alternative approach is needed to reach the optimal dependence of $\log(1/\eps)$. Since any $\log n$-degree $\F_2$ polynomial can be expressed as a $\sym\circ \AND_{\log n}$ circuit of size $n^{\log n}$, any improvement in the dependence of the seed length on $S$ would give nontrivial PRGs for $\log n$-degree polynomials, a breakthrough result. \ 

\begin{table}[t]

\renewcommand{\arraystretch}{1.6}
\centerline{
\begin{tabular}{|c|c|c|c|c|}
\hline
&  Circuit type &  Circuit size $S$ & Correlation bound & PRG seed length\\ \hline
\cite{v07} & $\{\sym,\thr\} \circ \ac$ & $n^{ c \log n}$ & $n^{-c_d \log n}$  & $2^{O(\sqrt{\log (S/\eps)})}$ \\ \hline 
\cite{ls11} & $\sym \circ \ac$ & $n^{c \log \log n}$ & $\exp(-n^{0.999})$   & $2^{O\big({\frac {\log S}{\log \log S}}\big)} + (\log (1/\eps))^{2.01}$ \\ \hline
\cite{ls11} & $\thr \circ \ac$ & $n^{c \log \log n}$ & $\exp(-n^{0.499})$   & $2^{O\big({\frac {\log S}{\log \log S}}\big)} + (\log (1/\eps))^{4.01}$\\ \hline 
\cite{st19} & $\{\sym,\thr\} \circ \ac_d$ & $n^{ c \log n}$ & $\exp(-\Omega(n^{0.499}))$     & $2^{O(\sqrt{\log S})} + (\log(1/\eps))^{4.01}$ 
\\ \hline \hline
{\bf This work} & $\{\sym,\thr\} \circ \ac$ & $n^{ c \log n}$ & $\exp(-\Omega(n^{0.999}))$     & $2^{O(\sqrt{\log S})} + (\log(1/\eps))^{2.01}$ 
\\ \hline
\end{tabular}
}
\caption{Correlation bounds against $\{\sym,\thr\} \circ \ac_d$ circuits and the PRGs that follow via the \cite{nw94} framework. In all previous work, the ``hard'' function used was the $\rw$ function, which was first considered by Razborov and Wigderson \cite{rw93}. Our work uses a better suited function.}
\end{table}

\subsection{Much Better PRGs Against Width-2 Branching Programs Reading $d$ Bits at a Time}

Usually, one constructs PRGs for natural \emph{computational} models, with the idea that we can drastically reduce the randomness we use if the randomized algorithm we are running can be simulated by such a model. low-degree polynomials is an extremely natural mathematical model with applications to circuit complexity, but some may not believe it is well grounded as a computational one and thus not worth finding a PRG for. However, the work of Bogdanov, Dvir, Verbin, and Yehudayoff \cite{bdvy13} showed the beautiful connection that PRGs for degree $d$ polynomials are also PRGs against a particular model described as \emph{width-2 length-$\poly(n)$ branching programs which read $d$ bits at a time}.

\begin{defn}[$(d,\ell, n)$-2BP (\cite{bdvy13}, adapted)]
    A $(d,\ell, n)$-2BP (or more colloquially a width-2 length-$\ell$ branching program over $n$ bits which reads $d$ bits at a time) is a layered directed acyclic graph, where there are $\ell$ layers and each layer contains two nodes, which we label by 0 and 1. Each vertex in each layer $j$ is associated with an arbitrary $d$-bit substring $x|_v$ of the input $x$. Each node in layer $j$ has $2^d$ outgoing edges to layer $j+1$ that are labeled by all possible values in $\bits^d$. On input $x$, the computation starts with the first node $v_{start}$ in the first layer, then follows the edge labeled by $x|_{v_{start}}$ onto the second layer, and so on until a node in the last layer is reached. The identity of this last node is the outcome of the computation.
\end{defn}

Such branching programs are a well-motivated computation model, which covers computation with only one bit of usable memory, low-degree polynomials, and small width DNFs. The survey of unconditional PRGs by Hatami and Hoza refers to this model as a compelling computational model that places low-degree polynomials in the computational landscape \cite{hh23}. \ 

Unfortunately, there is a ``$\log n$-degree barrier'' for PRGs and correlation bounds against low-degree polynomials. Current PRGs and correlation bounds are asymptotically tight for constant degree polynomials, but become \emph{trivial} at degree $\log n$, as can be seen by the current best known PRG for degree-$d$ polynomials by Viola which has seed length $O(d\log n + d2^d\log(n/\eps))$ \cite{vio08}. Obtaining nontrivial PRGs (or even correlation bounds) against $\log n$-degree polynomials has been a tantalizing open problem, and thus PRGs for $(d,\poly(n),n)$-2BPs also seemingly appeared to inherit this ``$d=\log n$ barrier'' due to the reduction result of \cite{bdvy13}. \ 

In this work, we construct PRGs against $(d,\poly(n),n)$-2BPs with exponentially better seed length, thereby giving nontrivial PRGs even in the regime $d = n^{1-o(1)}$. Define a $d$-junta to be a function $\phi:\bits^n\to\bits$ which is solely dependent on $d$ input bits (i.e. can be written as $\phi'(x_i)_{i\in S}$ for some subset $S\subset[n]$ of size $d$). To get our shortened seed length, we evade the $\log n$-degree barrier by instead showing the equivalence between PRGs for $(d,\poly(n),n)$-2BPs and PRGs for the XOR of $\poly(n)$ many $d$-juntas (denoted as $\junta^{\oplus \poly(n)}_{n,d}$). This class is already interesting in its own right, as it can be seen as a generalization of sparse $\F_2$-polynomials and combinatorial checkerboards (defined by Watson \cite{wat11} and also studied by Gopalan, Meka, Reingold and Zuckerman \cite{gmrz13}), as well as a specific class bounded collusion protocols studied by Chattopadhyay et al. \cite{cggklmz20}. However, we are not aware of any literature that studies $\junta^{\oplus m}_{n,d}$ specifically. \ 

 Our main technical contribution is strong correlation bounds for $\junta^{\oplus\poly(n)}_{n,d}$. In particular, we show the following. 
 
 \begin{thm}
     There exists an explicit function $f$ such that \[\corr(f,\junta_{n,d}^{\oplus\poly(n)})\le \exp\left(-\frac{n}{d2^{O(\sqrt{\log n})}}\right)\]
 \end{thm}

 By combining this with the ``hardness-to-randomness'' framework of Nisan and Wigderson \cite{nw94}, we construct a PRG of seed length $2^{O(\sqrt{\log n})}d^2\log^2(1/\eps)$. This is only a quadratic factor away from optimal dependence on $d$ and $\eps$. Improving the dependence on $n$ would be a breakthrough, since if we set $n' = 2^{\sqrt{\log n}}$, a $(d,n,\poly(n))$-2BP can simulate any $\log n'$-degree polynomial over $x_1,\dots x_{n'}$, and so having seed length $o(n')$ would effectively break the $\log n$-degree barrier for $\F_2$-polynomial PRGs.
 
 Interestingly enough, by combining a ``simplification under restriction'' approach pioneered by Ajtai and Wigderson \cite{aw85} with a PRG for sparse $\F_2$-polynomials by Servedio and Tan \cite{st22}, we can construct a PRG against $\junta^{\oplus\poly(n)}_{n,d}$, and thus $(d,\poly(n), n)$-2BPs, with seed length $d2^{O(\sqrt{\log(n/\eps)})}$. This gives us an optimal dependence on $d$, but an exponentially worse dependence on $\eps$. This suggests perhaps that with a combination of these two approaches, one might be able to achieve a seed length of $2^{O(\sqrt{\log n})}d\log(1/\eps)$.

\subsection{Near-Optimal Bounds Against High Degree Set-Multilinear Polynomials}

As explained earlier, a central open question in complexity theory is to establish better-than-$O(1/\sqrt{n})$ nontrivial correlation bounds against $\Omega(\log n)$-degree polynomials. In order to make progress on this question, it is natural to consider structured low-degree $\F_2$-polynomials. This is what the work of Bhrushundi et al.\ does \cite{bhhkk}.


Define a polynomial $p:\bits^n\to \bits$ as set-multilinear over a partition $X = (X_1,\dots, X_d)$ of the input bits if every monomial contains \emph{at most one} variable from each $X_i$ (this is slightly more general than the usual definition of \emph{exactly one}). The work of Bhrushundi et al.\ \cite{bhhkk} shows that a random degree $d$ set-multilinear tensor has exponentially small correlation against generic degree $d/2$ $\F_2$-polynomials for $d = \Omega(n)$. Toward making this correlation bound explicit, they defined $\ffm(X_1,\dots, X_d) = \lsb(X_1\cdot X_2\cdots X_d)$, where multiplication is done by treating $X_i$ as field elements, and $\lsb$ outputs the least significant bit of the string. Bhrushundi et al. were able to give exponentially small correlation bounds against polynomials up to degree $o(n/\log n)$ which are set-multilinear over the fixed partition $(X_1,\dots, X_d)$. 
However, this leaves much to be desired. The partition with respect to which the polynomial is set-multilinear over needing to be fixed and dependent on $\ffm_d$ feels like an extremely strong and asymmetric condition. Can we uncorrelate against degree $< d$ polynomials set-multilinear over \emph{any} equipartition of $X$ into $d$ parts? Can the parts be unequal? Can we have more than $d$ of them? \ 

We show affirmative answers to all the above questions. If we take $\delta > 0$ to be an arbitrarily small constant, we can obtain exponentially small correlation against degree $ < n^{\delta}$ polynomial for which there exists \emph{some} partition of $X$ into up to $n^{1-\delta}$ (not necessarily equal) parts such that $p$ is set-multilinear over it. Notice that improving $n^{1-\delta}$ parts to $n$ would be a breakthrough, since all polynomials are set-multilinear over the $n$-partition of $X = (x_1,\dots, x_n)$. 

To do so, we fortify the hard function $\ffm$ with an extractor. Let $\Ext(X,W)$ be a strong linear seeded extractor (for each fixing of $W$, $\Ext(\cdot, W)$ is linear). For some parameter $d$, define the function \[\extffm_d(X_1,\dots, X_d,W) := \lsb(\Ext(X_1,W)\cdot \Ext(X_2,W)\cdot\ldots\cdot \Ext(X_d,W)),\] where multiplication is done over a finite field, and $\lsb$ outputs the least significant bit of the string. First note that $\extffm_d(X_1,\dots, X_d, W)$, for a fixed $W$, is set-multilinear over $X_1,\dots, X_d$. Hence, our intuition that set-multilinear polynomials might correlate the most with the hard function is preserved in $\extffm$ as well. Using $\extffm$, we are able to obtain correlation bounds against the more intuitive notion of set-multilinear polynomials, where the structure of the partition does not matter. This gives more leeway since now if we want to implement this approach towards correlation bounds against low-degree polynomials, there is a larger class of set-multilinear polynomials to which we can reduce generic polynomials.

\section{Technical Overview Of the Results}

In this section, we give the overview of the proofs of the main results we covered above.

\subsection{Stronger Correlation Bounds Against $\{\sym,\thr\}\circ \ac$}
We focus on showing stronger correlation bounds against $\{\sym,\thr\}\circ\ac$, since the subsequent arguments turning this into PRGs against $\ac$ with a few $\{\sym,\thr\}$ gates are standard. The blueprint behind this argument follows the ``simplification under restrictions'' approach of previous works, but most similarly of Tan and Servedio \cite{st19}. A random restriction is a random partial assignment where for each variable, it is left unfixed (or ``alive'') with probability $p$, and is otherwise set to a uniform bit. \cite{st19} shows that under a random restriction, the hard function $\rw_{m,k,r}$ maintains integrity and uncorrelates with multiparty protocols, while $\{\sym,\thr\}\circ\ac$ simplifies to a short multiparty protocol. However, the roadblock met in \cite{st19} that prevents a correlation bound of $2^{-\Omega(n)}$ and only gives one of size $2^{-\Omega(n^{.499})}$ is due to the parameters in $\rw_{m,k,r}$ being in contention with each other. To elucidate, if $n$ is the input size, then we must have $mkr = n$. Via the analysis done in \cite{st19}, the correlation bound ends up being in the form of $2^{-\Omega(m)} + 2^{-\tilde{\Omega}(r)}$, which forces any established correlation bound to be at best $2^{-\Omega(\sqrt{n})}$. \ 

To understand why both conflicting terms appear, we give a quick overview of the argument of \cite{st19}. First, $\rw_{m,k,r}$ (as defined in \Cref{eqn:rw}) can be thought of as a fortified version of the \emph{generalized inner product}, $\gip_{m,k}(x_1,\dots, x_k) := \sum_{i=1}^m\prod_{j=1}^k x_{ij}$, where each variable is now replaced by the parity of $r$ new variables. This is effective against random restrictions, since as long as one of the $r$ copies $x_{ij1},\dots ,x_{ijr}$ survive the restriction, the corresponding term $x_{ij}$ in $\gip$ will survive. They argue that after applying a random restriction $\rho$, the $\{\sym,\thr\}\circ\ac$ circuit simplifies to a short multiparty protocol, while $\rw_{m,k,r}|_{\rho}$ is still capable of computing $\gip_{m/2,k}$ with high probability. Conditioning on this, previous results of Babai, Nisan, and Szegedy \cite{bns89} show that $\gip_{m/2,k}$ has $2^{-\Omega(m/2^k)}$ correlation against these multiparty protocols, which explains the emergence of the $2^{-\Omega(m)}$ term in the correlation. \ 
Conditioning on $\rw_{m,k,r}|_\rho$ being able to compute $\gip_{m/2,k}$ introduces an additive error to the correlation corresponding to the probability $\rw_{m,k,r}|_\rho$ \emph{fails} to simplify. \cite{st19} bounds this by the chance that all $r$ copies of some variable $x_{ij}$ becomes fixed by a restriction, which will be $(1-p)^r\approx \exp(-pr)$, explaining the occurrence of the $2^{-\Omega(r)}$ term in the correlation. 

In summary, the argument of \cite{st19} requires that $r$ be large to strongly fortify the hard function against random restrictions, while $m$ needs to be large to have a stronger correlation bound against multiparty protocols. However, with the constraint $mr\le n$, we are forced to compromise and reach the setting $m = r \approx \sqrt{n}$. \ 

We now propose an abstraction of the hard function, which naturally yields a stronger correlation bound. If we define $\oplus_{m,r}: (\bits^r)^m\to \bits^m$ to be \[\oplus_{m,r}(x_1,\dots, x_m) = \left(\sum_{i=1}^r x_{1i},\dots, \sum_{i=1}^r x_{mi}\right),\] we observe  $\rw_{m,k,r} = \gip_{m,k}\circ \oplus_{m,r}^k(x_1,\dots, x_k) := \gip_{m,k}(\oplus_{m,r}(x_{1}),\dots, \oplus_{m,r}(x_k))$. The key insight is that our argument can be generalized to not just $\rw$, but any function \[f\circ \Ext^k := f(\Ext(x_1),\dots, \Ext(x_k))\] where $f$ is average-case hard for multiparty protocols, and $\Ext$ is an \emph{oblivious bit-fixing source extractor} (OBF extractor). Informally, an oblivious bit-fixing source extractor for min-entropy $k$ is a function $\Ext$ such that if $\mathbf{X}$ is uniform over $\bits^n$ and $\rho$ is a restriction which leaves $\ge k$ bits alive, the output $\Ext({\mathbf X}|_\rho)$ is close to uniform. Recall our approach first applies a random restriction to simplify our circuit to a small multiparty protocol, which we then deal with using $\gip$. If the random restriction leaves sufficiently many variables alive with high probability, then $f\circ \Ext^k$ should still behave like $f$ due to $\Ext$ being an OBF extractor. Since the circuit is now a multiparty protocol, the average-case hardness of $f$ gives us a correlation bound (the argument is actually not as simple as this: see \Cref{sec:nuance} for an explanation).

Notice in the $\rw$ construction and the setting of parameters $m=r\approx \sqrt{n}$, $\oplus_{m,r}$ is an OBF extractor which maps $n$ bits to $\sqrt{n}$ bits. But this means that the input to the outer $\gip$ function will only have $\approx \sqrt{n}$ bits, so the best correlation bound we can hope to achieve is $\exp(-\Omega(\sqrt{n}))$. The restrictions used in the proof leave $n^{.99}$ variables alive with high probability, so intuitively we could hope that all these $n^{.99}$ ``bits of randomness'' could be preserved for $\gip$ (or in general any $f$) rather than only $\sqrt{n}$, potetially resulting in a $\exp(-\Omega(n^{.99}))$ correlation bound alive. We do just this by using a much better OBF extractor of Kamp and Zuckerman \cite{kz07}.  By making this intuition more formal using techniques developed by Viola and Wigderson \cite{vw07}, we obtain $2^{-\Omega(n^{1-O(1)})}$ correlation bound. The idea of replacing parities with better suited extractors has also appeared in previous work \cite{krt13,ckksz14,css16,kkl17}. However, as discussed earlier, they have only been used to boost lower bound arguments that solely use random restrictions. In this result, we are boosting a lower bound argument incorporating a communication complexity argument, which requires more care (see \cref{sec:nuance}).

\subsection{PRGs for $\junta^{\oplus t}_{n,d}$ and $(d,t,n)$-2BPs}

Our PRG construction blueprint can be briefly described as follows. We first establish correlation bounds against $\junta^{\oplus t}_{n,d}$. We then put this through the Nisan-Wigderson ``hardness vs. randomness'' framework to create a PRG against $\junta^{\oplus t}_{n,d}$. We then show that PRGs that fool $\junta^{\oplus t}_{n,d}$ actually fool $(d,t,n)$-2BPs, making the $\junta^{\oplus t}_{n,d}$ PRG our final construction. We first discuss why PRGs for $\junta^{\oplus t}_{n,d}$ imply PRGs for $(d,t,n)$-2BPs, and then discuss the techniques needed to show strong correlation bounds against $\junta^{\oplus t}_{n,d}$. 

\subsubsection{PRGs for $\junta^{\oplus t}_{n,d}$ $\implies$ PRGs for $(d,t,n)$-2BPs}
Adopting the exposition in \cite{hh23}, the previous work of \cite{bdvy13} can be outlined as follows. Consider a $(d,t,n)$-2BP $B$. Observing that all transition functions in $B$ are $d$-juntas, one can derive that $B(x) = B'(\phi_1,\dots, \phi_{2t}(x))$, where $B'$ is a $(1,t, 2t)$ branching program. By Fourier expanding $B'$, this can be decomposed as \[B(x) = \sum_{S\subset[2t]}\widehat{B}(S)(-1)^{\sum_{i\in S}\phi_i(x)}.\] \cite{bdvy13} shows that $\sum_{S\subset[t]} |\widehat{B}(S)|$ is bounded, so by linearity of expectation and the Triangle Inequality, it suffices to fool the terms $(-1)^{\sum_{i\in S}\phi_i(x)}$. The approach in \cite{bdvy13} makes the observation that each $\phi_i$, by virtue of being a $d$-junta, can be written as a degree $d$ polynomial. Consequently, a PRG for degree $d$ polynomials will fool $(d,t,n)$-2BPs with seed length $O(d\log n + d2^d\log(n/\eps))$. The issue here is that at $d = \log n$, the seed length becomes trivial. \ 

However, we can notice that the $\F_2$-polynomial $p(x)\coloneqq \sum_{i\in S}\phi_i(x)$ has some additional structure. If $t = \poly(n)$, $p$ is the sum of only a polynomial number of $d$-juntas. If there was a way to leverage this, and get a better PRG that fools $\junta^{\oplus t}_{n,d}$, then we might hope to get nontrivial PRGs even in the regime $d = \Omega(\log n)$.

This observation already yields nontrivial PRGs for $d = \omega(\log n)$. Servedio and Tan \cite{st19} provide a PRG fooling $\F_2$-polynomials with $S$ terms with seed length $2^{O(\sqrt{\log S})}\log(1/\eps)$. Since each junta can be written as a polynomial with up to $2^d$ terms, each $g\in \junta_{n,d}^{\poly(n)}$ can be written as a polynomial with $S = 2^d\poly(n)$ terms, yielding a PRG with seed length $(2^{O(\sqrt{d})} + O(\log n))\log(1/\eps)$. Hence we get nontrivial seed length for $d = o(\log^2 n)$ \footnote{it is actually the case that a PRG from \cite{lvw93} already gets nontrivial seed length in the same regime, albeit with exponentially worse dependence in $\eps$}. However, we proceed alternatively to get an exponentially better seed length.

\subsection{The Nisan-Wigderson Framework and Correlation Bounds for $\junta_{n,d}^{\oplus \poly(n)}$}
We will once again use $f\circ\Ext^k$ \footnote{we also precompose with parities in the formal argument} as our hard function to establish exponentially small correlation bounds against the class, and then apply the Nisan-Wigderson \cite{nw94} framework to construct the PRG. The latter portion is straightforward, so we focus on establishing the correlation bounds. \

Let $g\in\junta_{n,d}^{\oplus \poly(n)}$. We first show that there exists a subset of variables, $S$, such that when arbitrarily fixing bits outside of this set, $g$ can be expressed as a sparse $\F_2$ polynomial, while each input block of $f\circ\Ext^k$ heavily intersects $S$. Hence if we fix $X_{\bar{S}}$ and take the correlation over $S$, each input block still maintains high min-entropy while $g$ becomes a sparse polynomial, which is a small $\sym\circ \ac$ circuit. Since the hard function is also the same, we can then apply techniques in the previous section to conclude.

\subsection{Correlation Bounds against Set-Multilinear Polynomials}

Recall that \cite{bhhkk} has shown $\ffm_d$ uncorrelates against any lower degree polynomial which is set-multilinear over $(X_1,\dots, X_d)$. The key ingredient behind proving strong correlation bounds against set-multilinear polynomials over arbitrary parititons is to first fortify each input block with extractors, and instead consider $\extffm_d$. This allows us to establish the following structural lemma, which intuitively states that even if you do not start out with a polynomial that is set-multilinear over $(X_1,\dots, X_d)$, if not too many bits in each input block can be restricted to 1s such that the resulting function is set-multilinear over $(X_1,\dots, X_d)$ induced by the live variables in each block, exponential correlation bounds can still be obtained. 

\begin{thm}
Let $g$ be a polynomial of degree $< d$. Let $S_1,\dots, S_{d}\subset [n/d]$ be subsets, and let $\rho$ denote the restriction created by fixing the bits in $X_i$ whose index is outside $S_i$ to $1$ for each $i\in[d]$. If the restricted function $g|_\rho(X_1,\dots, X_d)$ becomes set-multilinear in $(X_1,\dots, X_d)$, then have $$ \corr(\extffm_d,g)\le 2^{-\Omega(\frac{n}{cd})}.$$    
\end{thm}

To explain the proof at a high level, if the sets $S_i$ we leave alive aren't too small, then our strong extractor (conditioned on a good seed) will keep each block $\Ext(X_i,W)$ approximately uniform, and since the restricted function $g|_\rho$ is now set-multilinear over $(X_1,\dots, X_d)$ we may use a similar approach as \cite{bhhkk} to prove the theorem.

It turns out that through a combinatorial argument, one can show that polynomials which are set-multilinear over a large number of blocks can be turned into polynomials set-multilinear over $(X_1,\dots, X_d)$ by fixing not too many bits per input block $X_i$. The correlation bounds then follow from the structural lemma.

\subsection{A Technical Nuance Regarding Extractor Fortification}
\label{sec:nuance}

Say we are given a class $\calC$ that simplifies under restrictions proportionally to how aggressive the restriction is. For example, say with high probability under a mild random restriction $\rho_1$, $C|_{\rho_1}$ simplifies to a class $\calC_1\subset \calC$, and with high probability under a more aggressive random restriction $\rho_1\circ\rho_2$, $\calC|_{\rho_1\circ \rho_2}$ simplifies to a much smaller class $\calC_2\subset \calC_1$. A common technique to prove correlation bounds is to take a hard function $f:\bits^n\to\bits$ that remains hard after a random restriction. For example, if we know $\E_{\rho_1}\corr(f|_{\rho_1}, \calC_1) = \eps_1$, we can bound \[\corr(f,\calC) \le  \E_{\rho_1}\corr(f|_{\rho_1},\calC|_{\rho_1})\approx \E_{\rho_1}\corr(f|_{\rho_1}, \calC_1)= \eps_1, \] where $\approx$ hides the probability $\calC$ does not simplify. With this, one might hypothesize that if we precompose with an OBF extractor $\Ext:\bits^m\to\bits^n$, one must have $\corr(f\circ\Ext,\calC)\ll \eps_1$.

In particular, if $\calC_2$ is much simpler than $\calC_1$, we would expect $\eps_2\coloneqq \corr(f|_{\rho_1},\calC_2) \ll \corr(f|_{\rho_1},\calC_2) = \eps_1$. Then, by the extractor property, we would have $f\circ\Ext|_{\rho_2}(U)\approx f(U')$, where $U$ and $U'$ are uniform strings (but could be correlated depending on the live variables of $\rho_2$). Hence, \begin{align}\corr(f\circ \Ext|_{\rho_2}, \calC|_{\rho_2}) = |\E(-1)^{f\circ\Ext|_{\rho_2}(U) + g|_{\rho_2}(U)}|\approx |\E(-1)^{f(U') + g|_{\rho_2}(U)}|.\label{eqn:heuristic-bound}\end{align} If $U'$ was almost always a copy of $U$ (over the randomness of $\rho_2$), the above would let us deduce $\corr(f\circ\Ext|_{\rho_2},\calC_{\rho_2})\approx \corr(f,\calC_{\rho_2})$. This allows us to deduce a much stronger correlation bound as follows.
\begin{align*}
\corr(f\circ \Ext,\calC) \le \E_{\rho_2}\corr(f\circ \Ext|_{\rho_2}, \calC|_{\rho_2})& \approx  \E_{\rho_2}\corr(f, \calC|_{\rho_2}) \\ &  \le \E_{\rho_1,\rho_2}\corr(f|_{\rho_1}, \calC|_{\rho_1\circ\rho_2}) \\ & \approx \corr(f|_{\rho_1},\calC_2) = \eps_2 .\end{align*}

Of course, we cannot assume something so strong. It is possible that $(U',U)$ are adversarially correlated such that the expression in \Cref{eqn:heuristic-bound} is extremely large. In this paper, we find two ways to circumvent this obstacle, resulting in three new correlation bounds.



Since the correlation between $U$ and $U'$ is the issue, we might hope to bypass this obstacle by bounding the RHS of \Cref{eqn:heuristic-bound} by an expression independent of $g|_\rho(U)$. This is indeed possible for multiparty protocols, as Viola and Wigderson \cite{vw07} showed the correlation of a function $f$ against $k$-party protocols is bounded by the ``$k$-party'' norm of $f$. Our correlation bounds for $\{\sym,\thr\}\circ\ac$ crucially rely on this to make the extractor fortification argument work. 

Another way one might imagine a fortification lemma might hold is if $\Ext|_\rho$ was ``easy to invert''. To see what we mean, lets say for simplicity that $\Ext|_\rho$ was a bijection and had inverse $\Ext^{-1}$. Then we could see by a change of variables \[|\E(-1)^{f\circ\Ext|_\rho(U) + g|_\rho(U)}|\approx |\E(-1)^{f(U) + g|_\rho(\Ext^{-1}(U))}|.\] Hence if $\Ext^{-1}$ was computationally simple, we might be able to reason that $g|_\rho\circ\Ext^{-1}$ is also very simple, and thus has very small correlation against $f$. This intuition is what drives the analysis in the correlation bounds against set-multilinear polynomials. We use a linear seeded extractor, which makes the maps linear and thus ``easy to invert''. Furthermore, this inversion reasoning explains why for our correlation bounds involving multiparty protocols, we need to apply an independent extractor to each party's input rather than one extractor for the whole input. If we did the latter, then the inverse $\Ext^{-1}$ would add correlation among the $k$ input strings in $g|_\rho\circ\Ext^{-1}$, thereby \emph{increasing} the communication complexity of this function.

For correlation bounds against width-2 branching programs, we use both phenomena, as we precompose with an OBF extractor, and then again with a simple parity extractor. 

It is an interesting question to further explore the connection between extractors and correlation bounds.

\section{Preliminaries}

For positive integer $n$, $[n]:= \{1,\dots, n\}$ and $\binom{[n]}s$ is the set of all subsets of $[n]$ with $|S| = s$. We denote $e(x):= (-1)^x$.

\subsection{Convention About Input Blocks}
\label{sec:blockconv}
We will canonically fix a partition of bit strings into $d$ contiguous blocks, each with $n/d$ bits. In particular, any $X\in\bits^n$ can be written as $X = (X_1,\dots, X_d)$ where each $X_i$ is the $n/d$-bit substring. If a string $Y\in\bits$ is defined, $Y_i$ will be assumed to mean the length $n/d$ substring of $Y$ contained in the $i$th input block, defined with respect to the canonical partition. Also, we will denote $X_{-i}:=(X_1,\dots, X_{i-1},X_{i+1},\dots, X_d)$ to be the input with the $i$th block removed. 

For a string $X\in \bits$, we may sometimes identify the $n/d$ bit string $X_i$ as an $n$-bit string in the following way: the $i$th block is filled with $X$, and all other blocks are filled with 0s. Hence, if we interpret bit strings as elements of $\F_2^n$, and we have $X,Y\in \F_2^n$, the expression $X+Y_i$ is well defined.

For parameters $k, d\le n$ and two functions $f:(\bits^m)^k\to\bits$ and $g:\bits^{n/k}\to \bits^d$, we will define \[f\circ g^k = f(g(X_1),\dots, g(X_d))\]
\subsection{Finite Fields}
\label{sec:finitefield}

We will be working with finite fields of characteristic 2. For the finite field over $2^n$ elements, $\F_{2^n}$, we can naturally identify each element with an $n$-bit string. It is well known $\F_{2^n}$ can be constructed by taking a degree $n$ polynomial $E(x)$ which is irreducible over $\F_2[x]$, and considering the quotient $\F_2[x]/E(x)$. \ 

Consequently, via polynomial division by $E(x)$, each element $a\in \F_{2^n}$ can be uniquely identified with a degree $n-1$ $\F_2[x]$ polynomial, say $\sum_{i=0}^{n-1}c_ix^i$. We can then define the corresponding $n$-bit string representation of $x$ to be $(c_{n-1},c_{n-2},\dots, c_0)$ (noting that the least significant bit is the constant term). Furthermore, if we explicitly write out a string in $\F_{2^n}$ with less than $n$ bits, simply pad all the leading digits with 0s to turn it into an $n$-bit string. (e.g. $11$ is interpreted the same as $0^{n-2}11$).\ 

With this polynomial-to-string interpretation, addition and multiplication become well-defined. We note that for $x,y\in \F_{2^n}$, $x+y$ behaves exactly like adding two $\F_2^n$ vectors. We can now define the multiplication $x\cdot y$ by converting them to polynomials, taking the product, modding out by $E(x)$, and then writing the coefficients as a bit string. We will also define $\langle x, y\rangle = \sum_{i=1}^n x_iy_i$ where the RHS arithmetic is over $\F_2$.

\begin{defn}[character]
A map $\chi:\F_{2^n}\to\F_2$ is called an \emph{additive character} if for all $x,y\in \F_{2^n}$, $\chi(x+y) = \chi(x) + \chi(y)$. It is nontrivial if it is not the zero function.
\end{defn} 
Since $\F_{2^n}$ is an $n$-dimensional vector space, we see the valuations on $n$ basis vectors uniquely define the character. Consequently there are $2^n$ such characters. Notice we can conveniently characterize all characters either by $\chi_c(x) = \langle x, c\rangle$, or by fixing some character $\chi$, and then defining $\chi_c(x) := \chi(c\cdot x)$. This can be seen by verifying these maps are characters, are distinct, and that there are $2^n$ of them (the latter is obvious since there are $2^n$ values of $c$).

\subsection{Models of Computation}

\begin{defn}[$\F_2$-polynomials]
    An $\F_2$-polynomial (or polynomial for short) is a function of the form $p(x) := \sum_{S\subset[n]}c_S \prod_{i\in S}x_i$ for some $c_i\in \F_2$ (all arithmetic here are over $\F_2$).
\end{defn}

\begin{defn}[set-multilinearity]
    An $\F_2$-polynomial $p$ is \emph{set-multilinear} over a partition $(X_1,\dots,X_d)$ of variables if every monomial of $p$ contains at most one variable from each $X_i$. Notice that all polynomials are trivially set-multililinear over $(x_1,\dots, x_n)$.
\end{defn}

\begin{defn}[junta]
Define the class $\junta_{n,k}$ to be a function $\phi:\bits^n\to\bits$ which is solely dependent on $k$ input bits (i.e. can be written as $\phi'(x_i)_{i\in S}$ for some subset $S\subset[n]$ of size $k$). Define $\junta_{n,k}^{\oplus t}$ to be the class of functions which is the parity of $t$ $k$-juntas.
\end{defn}

\begin{defn}[$k$-party NOF protocol]
A boolean function $f:(\bits^{n/d})^d$ can be computed by a $k$-party NOF protocol with $c$ bits of communication if on input $X = (X_1,\dots, X_d)$, $d$ players, can take turns writing a bit on the board, where player $i$'s bit can only depend on $X_{-d}$ and the other bits on the board, and the $c$th bit written is $f(X)$. We denote this class of functions to be $\Pi_k^c$
\end{defn}

\subsubsection*{Circuits}
 We measure the size of a circuit by the total number of wires (including input wires) in it. $\ac_d$ are depth $d$ circuits with unbounded fan-in whose gate set is $\{\AND, \OR, \NOT\}$. $\sym$ is a gate which computes an arbitrary symmetric function, and $\thr$ is a gate which computes an arbitrary linear threshold function. In general, if we have a gate $G$, a subscript $G_k$ will refer to its fan-in (in this case, $G$ is fixed to have fan-in $k$).

\begin{defn}[$(d,\calC)$-tree]
    Let $d$ be an integer and $\calC$ a computational model (e.g. a circuit class). A function is computable by a $(d,\calC)$-tree if it is computable by a depth $t$ decision tree with $\calC$ functions as its leaves. That is, there exists a depth $d$ decision tree $T$ such that for every path $\pi$ in $T$, $F|_\pi\in \calC$.
\end{defn}

\subsection{Probability}

We will denote $U_m$ to be the uniform distribution over the finite set $\bits^m$. We will also denote $S\subset_p T$ to be a random subset of $T$ where each $t\in T$ is added to $S$ independently with probability $p$.

\begin{defn}[$k$-wise uniform]
    Consider a distribution $D$ over $(\bits^{n/d})^d$. We say that $D$ is $k$-wise uniform if for all subsets $S=\{i_1,\dots, i_k\}\subset [d]$ and all strings $y_1,\dots, y_k\in\bits^{n/d}$, \[\Pr_{X\sim D}[\forall j, X_{i_j} = y_j] = 2^{-kn/d}.\]
\end{defn}

\begin{defn}[$\eps$-close in distribution]
Let $D_1$ and $D_2$ be distributions over $\bits^n$. We say $D_1\approx_{\eps} D_2$, or equivalently $D_1$ is $\eps$-close to $D_2$, if for all $S\subset\bits^n$, \[|\Pr_{x\sim D_1}[x\in S] - \Pr_{x\sim D_2}[x\in S]|\le \eps.\]
    
\end{defn}

\subsection{Random Restrictions and Partial Assignments}
A \emph{partial assignment} or \emph{restriction} is a string $\rho\in \{0,1,\star\}^n$. Intuitively, a $\star$ represents an index that is still "alive" and hasn't been fixed to a value yet.

We also define a composition operation on partial assignments. For two restrictions $\rho^1,\rho^2$, define $\rho^1\circ \rho^2$ so that \[(\rho^1\circ\rho^2)_i =\begin{cases} \rho^1_i & \rho^1_i\neq \star \\ \rho^2_i & \rho_i^1 = \star.\end{cases} \] Intuitively, one can see this as fixing bits determined by $\rho^1$ first, and then out of the remaining alive positions, fix them according to $\rho^2$.\

A \emph{random restriction} is simply a distribution over restrictions. A common random restriction we will use is $R_p$, the distribution where each index will be assigned $\star$ with probability $p$, and $0,1$ each with probability $\frac{1-p}2$.

The main reason for defining restrictions is to observe their action on functions. Given a restriction $\rho$ and function $f:\bits^n\to \bits$, we define $f|_{\rho}: \bits^n\to\bits$ to be the restricted function mapping $f|_{\rho}(x) := f(\rho\circ x)$. 

\subsection{Pseudorandomness}

Our work will involve working with pseudorandomness primitives, like pseudorandom generators (PRGs) and randomness extractors (or simply extractors).

\begin{defn}[$\eps$-PRG]
A polytime computable function $G:\bits^s\to\bits$ is an $\eps$-PRG for a subset $\calF$ of functions $\bits^n\to\bits$ if for all $f\in\calF$, \[|\E_{x\sim U_n}[(-1)^{f(x)}]-\E_{s\sim U_s}[(-1)^{f(G(s))}]|\le \eps.\] We also say that $G$ $\eps$-fools $\calF$. The parameter $s$ is the seed length. In this paper, we will use a PRG of \cite{st22} which $\eps$-fools $\F_2$ polynomials with $\le S$ terms with seed length $2^{O(\sqrt{\log S})}\log(1/eps)$.
    
\end{defn}

\begin{defn}[min-entropy]
    Let $D$ be a distribution over $\bits^n$, and define $\text{supp}(D) = \{y\in\bits^n: \Pr_{x\sim D}[x=y] > 0\}$. Define the min-entropy of $D$ to be the quantity \[-\log\left(\max_{x\in\bits^n}\Pr_{y\sim D}[y=x]\right).\] It is helpful to note that if for a particular $k$ and all $y\in \bits^n$, all probabilities $\Pr_{x\sim D}[x=y]\le 2^{-k}$, then we know $D$ has min-entropy $\ge k$.
\end{defn}

\begin{defn}[Strong/Linear/Seeded Extractors] A $(k,\eps)$-seeded extractor is a function $\Ext:\bits^n\times\bits^d\to\bits^m$ such that for any $D$ with min-entropy $\ge k$, we have for $\textbf{X}\sim D$ and $\textbf{W}\sim U_d$ the following  \[\Ext(\textbf{X},\textbf{W})\approx_\eps U_m.\] $\Ext$ is a strong seeded extractor if we also have \[\Pr_{w\sim U_d}[\Ext(\textbf{X},w)\approx_\eps U_m]\ge 1-\eps\] $\Ext$ is a linear seeded extractor if for every fixed $W$, $\Ext(\cdot, W)$ is linear over $\F_2$. The Leftover Hash Lemma \cite{ill89} allows us to construct a strong seeded $(k,\eps)$ extractor with seed length $2n$, $\Ext: \bits^n\cdot \bits^{2n}\to \bits^{k-2\log(1/\eps)}. $ 
    
\end{defn}

\begin{defn}[Oblivious Bit-Fixing Source Extractors]
 An $(n,k)$ oblivious bit-fixing source (or OBF) is a distribution $D$ over $\bits^n$ created by fixing some $n-k$ of the bits, and then filling in the remaining $k$ indices with uniform and independent bits. An $(k,\eps)$ oblivious bit-fixing source extractor (or OBF extractor) is a function $\Ext:\bits^n\to\bits^m$ such that for every $(n,k)$ OBF $D$, we have that for $\textbf{X}\sim D$, \[\Ext(\textbf{X})\approx_\eps U_m.\] For any $k>\sqrt{n}$, Kamp and Zuckerman \cite{kz07} allows us to construct $(k,2^{-\Omega(k^2/n)})$ OBF extractors $\Ext:\bits^n\to\bits^{\Omega(k^2/n)}$.
\end{defn}

\subsection{Correlation Bounds}

We will need some tools and definitions from the literature of correlation bounds. We first give a formal definition of correlation.

\begin{defn}[correlation]
    For two Boolean functions $f,g:\bits^n\to\bits$, and a distribution $D$ over $\bits^n$, define the correlation of $f$ and $g$ over $D$ to be \[\corr_D(f,g) = |\E_{x\sim D}(-1)^{f(x)+g(x)}|.\] If no distribution is mentioned, we always assume $D = U_n$. Furthermore, for a subset of functions $\calC$, we define \[\corr_D(f,\calC) = \max_{g\in\calC}\corr_D(f,g).\]
\end{defn}

Viola and Wigderson defined a convenient quantity $R_k$, which is very useful in bounding correlations against NOF protocols.

\begin{defn}[$k$-party Norm]
For a function $f:(\bits^{n/k})^k\to\bits$, define the $k$-party norm of $f$ to be \[R_k(f) := \E_{X_1^{(0)},\dots,X_k^{(0)},X_1^{(1)},\dots, X_k^{(1)}\sim U_{n/k}}e\left(\sum_{\delta\in\bits^k} f(X_1^{(\delta_1)},\dots, X_k^{(\delta_k)})\right).\]

\end{defn}

This norm is useful due to the following theorem.

\begin{thm}[\cite{vw07}]
\label{thm:vwcomm}
    Let $f:\bits^n\to\bits$ be arbitrary, and let $g$ be computable by a $d$-party NOF protocol exchanging $c$ bits. Then \[R_d(f)\le \corr(f,g)\le 2^c R_d(f)^{1/2^d}.\] 
\end{thm}

We will also use the following theorem of Nisan and Wigderson, which allow us to translate correlation bounds into PRGs.This version is seen in the survey of Hatami and Hoza \cite{hh23}

\begin{thm}[\cite{nw94}, \cite{hh23} Theorem 4.2.2]

\label{thm:nw}
Let $f:\bits^n\to \bits$. Suppose $h:\bits^r\to \bits$ is $\eps$-hard for $f\circ\junta_{r,k}$ with respect to the uniform distribution. Then there exists a PRG for $f$ with seed length $s = O(n^{\frac{1}{k+1}}\cdot r^2/k)$ and error $\eps n$.

\end{thm}

\section{Nearly Optimal Correlation Bounds against $\{\sym, \thr\}\circ \ac$}
\label{sec:fatsection}

We strictly improve upon the result \cite{st19} by proving a stronger correlation bound against $\{\sym, \thr\}\circ \ac$ circuits. This immediately gives PRGs against this class with improved seed length via the ``hardness vs. randomness'' framework \cite{nw94} All previous work \cite{v07,ls11,st19} looked at the function introduced in \cite{rw93} created by taking the generalized inner product of parities. We present a new function comprised of field multiplication of extractors in order to prove stronger correlation bounds. Let $m,n$ be parameters, and define $k:=n/d$.   We now prove the following result: 

\begin{thm}
\label{thm:circuit} 
Let $\Ext:\bits^k\to \bits^{.2k^{.996}}$ be a $(k^{.998}.2^{-.4k^{.996}})$ OBF-source extractor (explicit ones exist due to \cite{kz07}). Let $f:(\{0,1\}^{.2k^{.996}})^d\to \{0,1\}$ be any function such that $\corr(f,\Pi_d^d)\le 2^{-\Omega(k^{.996}/2^d)}$. Define $f\circ\Ext^d:(\{0,1\}^k)^d\to \{0,1\}$ to be the function $$f\circ\Ext^d(X) := f(\Ext(X_1),\dots,\Ext(X_d)).$$
Let $g$ be any function implementable by a $n^{O( \log n)}$-size $\{\sym, \thr\}\circ \ac$ circuit, and let $m = .0005\log n$. Then $$ \corr(f\circ \Ext^{m+1}, g)\le 2^{-\Omega(n^{.995})}.$$ 
\end{thm}

 In particular, by instantiating this template, say, with $\Ext$ being the extractor of \cite{kz07} and $f$ being either $\gip$ \cite{bns89} or $\ffm$ \cite{fg13}, we get explicit $f\circ\Ext^{m+1}$. We also note by simple adjusting of constants, we can get any $2^{-\Omega(n^{1-\eps})}$ for constant $\eps > 0$. This gives an improvement of the correlation bound given in \cite{st19} of $2^{-\Omega(n^{.499})}$.  

\begin{proof}
We follow the same approach as done in \cite{st19}. The uniform distribution can be expressed as applying a random restriction, and then filling in the remaining bits uniformly. For good random restrictions, we argue that $g$ simplifies to a $\{\sym,\thr\}\circ\AND_m$ circuit. We then argue that even after the random restriction, $f\circ \Ext^{m+1}$ maintains its structural integrity due to the extractor. We then finish the argument by using Hastad and Goldmann's connection between $\{\sym,\thr\}\circ\AND_m$ and NOF protocols, and the fact that $f$ has small correlation with $(m+1)$-party protocols.\\

The proof for the simplification of $g$ is the same as seen in \cite{st19} so we merely cite it here. The only change is the tuning of parameters. Here is the lemma restated for our use.

\begin{lem}
\label{lem:circtodt}
Let $g\in \{\sym,\thr\}\circ \ac_d$ with circuit size $s = n^{\tau\log n}$. Then for $p = \frac{1}{48}(48\log s)^{-(d-1)} $ \begin{align*}\Pr_{\rho\gets{\cal R}_p}[g|\rho &\text{ is not computed by $(.001pk,\{\sym_{s^2},\thr_{s^2}\}\circ \AND_{\log s}\})$-tree}] \\ & \le s\cdot 2^{-.001pk/2^d} \\ & = 2^{-\Omega_d(pk)}\end{align*}
\end{lem}

Notice that for constant $d$ this gives a bound of $2^{-\Omega(n/\polylog(n))}$, versus its use in \cite{st19} in which a $2^{-\Omega(\sqrt{n/\log n})}$ error was gained. We will see later that we can liberally set parameters here because our hard function maintains integrity even after traversing down a path of size $n/\polylog(n)$ (equivalent to randomly fixing $n/\polylog(n)$ bits), whereas the previous GIP function could only withstand $\sqrt{n}$ bits. This is result of using an OBF extractor with much better parameters than simply taking the XOR of many copies.\\

The leaves of our tree is now much simpler class of circuits, but it is not simple enough. Our correlation bounds can only handle circuits with fan in $m = O(\log n)$, but we currently have fan in $\log s  = O(\log^2 n)$. Fix a leaf $\ell$ of the tree, and let $\{C_1,\dots,C_{s^2}\}$ be a collection of subsets of $[n]$ where $C_i$ contains the  $\le \log s$ indices of the variables that feed into the $i$th $\AND_{\log s}$ gate in the bottom layer. We now use the following basic fact, as in \cite{ls11} and \cite{st19}, that there is a large subset of variables that minimally intersect with each $C_i$. 
\begin{claim}
\label{claim:qset}
A random ${\bf L}\subset_q[n]$ (add each element to ${\bf L}$ with probability $q$) satisfies $$\Pr[\exists i\in [s^2]\text{ such that }|C_i\cap {\bf L}| > m] \le s^2\binom{w}m q^m.$$\end{claim} Instantiating this claim with our parameter setting of $m$ and $s$, and setting $q = \Theta(n^{-.001})$ tells us $$\Pr[\exists i\in [s^2]\text{ such that }|C_i\cap {\bf L}| > m] \le \frac{1}s.$$
Hence there exists such an $L = L(\ell)$ such that restricting all bits outside $L$ makes only $\le m$ variables feed into each $\AND$ gate as desired.\\

To summarize, our restriction $\rho$ is sampled by a distribution $D$ specified by these three steps. \begin{enumerate}
\item We first perform restriction ${\cal R}_p$, 
\item and then randomly restrict $\le .001pk$ while walking down the depth-$.001pk$ tree to a leaf $\ell$,
\item and then randomly restrict all the variables alive in this leaf that is \emph{not} in the $L(\ell)$ set that we showed existed
\end{enumerate} 

At the end of this process, we have by the union bound that with all but $2^{-\Omega(-pk)}$ probability, $g|_\rho$ becomes a $\{\sym_{s^2},\thr_{s^2}\}\circ \AND_m$ circuit.\\

We now observe what happens to $f\circ\Ext^{m+1}$ under this restriction $\rho$. We claim $f\circ\Ext^{m+1}$ retains its structure. Our wish is for at least $k^{.998}$ bits in each block to survive. That way, we will have a high entropy oblivious bit-fixing source fed into each extractor, and the function will be able to continue to strongly uncorrelate with $m$-party protocols. In Step 1, we draw a restriction from ${\cal R}_p$. Notice the live variables are distributed like a set $S\subset_p [n]$. We see that by a simple Chernoff and union bound, $$\Pr_{{\bf S}\gets {\cal R}_p}\left[\exists i\in [m+1]\text{ such that } |X_i\cap {\bf S}| < \frac{pk}2\right]\le (m+1)2^{-\Omega(pk)}$$ 

Hence except for probability $m 2^{-\Omega(pk)} = 2^{-\Omega(n^{1-o(1)})}$, each block $X_i$ will have $\ge pk/2$ live variables. Conditioned on this, when we follow Step 2 and perform a random walk down the decision tree to a leaf, we will assign at most $.001pk$ bits, so we are guaranteed that each block $X_i$ will contain at least $.499pk$ live variables. Step 3 is to take set $L(\ell)$ and arbitrarily restrict variables outside of it. We showed there exists an $L(\ell)$ which minimally overlaps with the input variables to the $\AND_{\log s}$ gates, but we want it to simultaneously overlap heavily with each block. That way most of the $X_i$ will stay alive after restricting the bits outside of $L(\ell)$ The existence of such an $L(\ell)$ can be established by  ``completing the probabilistic method'' started a few paragraphs above. Conditioning on good restrictions so far, let $Y_i$ denote the variables that survived in $X_i$ (hence $|Y_i|\ge .499pk$).  We see that $$\Pr_{{\bf L}\subset_q [n]}\left[\exists i\in [m+1]\text{ such that }|Y_i\cap {\bf L}| < \frac{.499pqk}2\right] \le (m+1)2^{-\Omega(pqk)}.$$ Hence, the probability that ${\bf L}$ either intersects some $C_i$ too much or some $Y_i$ too little will happen with probability $\le \frac{1}s  + (m+1)2^{-\Omega(pqk)} \ll 1$. Thus there exists an $L(\ell)$ such that restricting all variables outside of it will simultaneously simplify $g$ to a $\{\sym_{s^2},\thr_{s^2}\}\circ \AND_m$ and also leave at least $\frac{.499pqk}2\ge .249 k^{.999}/\polylog(n)\gg k^{.998}$ variables alive. Stringing all three steps together, we know that except with probability $2^{-\Omega(-pk)}$, our random restriction $\rho$ reduces $g$ to $\{\sym_{s^2},\thr_{s^2}\}\circ \AND_m$, while simultaneously keeping $\ge k^{.998}$ variables in each $X_i$ block alive.\\

We are now in the final phase of the argument where we now directly bound the correlation against the simplified circuit. We first state the results that will convert our circuits to NOF protocols.

\begin{thm}[\cite{hg91}]
\label{thm:hg91}
 Let $f : \{0, 1\}^n\to  \{0, 1\}$ be a Boolean function computed by a size-$s$ $\sym\circ\AND_m$
circuit. Then for any partition of the $n$ inputs of $f$ into $m + 1$ blocks, there is a deterministic NOF
$(m + 1)$-party communication protocol that computes $f$ using $O(m \log s)$ bits of communication.
\end{thm} 

\begin{thm}[\cite{nis93}]
\label{thm:nis93}
 Let $f : \{0, 1\}^n \to \{0, 1\}$ be a Boolean function computed by a $\thr\circ\AND_m$
circuit. Then for any partition of the $n$ inputs of $f$ into $m + 1$ blocks, there is a randomized NOF
$(m + 1)$-party communication protocol that computes $f$ with error $\gamma_{err}$ using $O(m^3\log n \log(n/\gamma_{err}))$
bits of communication.
\end{thm}

We now need to show an average-case hardness result for $f\circ \Ext^{m+1}|_\rho$ against NOF protocols. To do so, we will first calculate the $k$-party norm of $f\circ\Ext^{m+1}|_{\rho}$.

\begin{lem}
\label{lem:extsavesnorm}
Let $\rho$ be a restriction which keeps $\ge k^{.998}$ variables in each $X_i$ alive. Then $R_{m+1}(f\circ\Ext^{m+1}|\rho)\le R_{m+1}(f) + 4(m+1)\cdot 2^{-4k^{.996}}$
\end{lem}

\begin{proof}
     Now notice that \begin{align}
    R_{m+1}(f\circ\Ext^{m+1}|_\rho) = \E_{X^{(0)},X^{(1)}}e\left(\sum_{\delta\in\bits^{m+1}} f(\Ext(X_1^{(\delta_1)}|_\rho ),\dots, \Ext(X_{m+1}^{(\delta_{m+1})}|_\rho))\right)\label{eqn:kpartyerror}
    \end{align}
 By assumption of $\rho$, each $X_i^{(\delta_i)}|\rho$ over uniform $X_i$ is an OBF source with min-entropy $k^{.998}$, and so each $\Ext|_\rho(X_i)\approx_{2^{-4k^{.996}}} U_{.2k^{.996}}$. Since all $X_i^{(b)}$ for $i\in[m+1], b\in\bits$ are mutually independent, it follows by a hybrid argument that  \[(\Ext|_\rho(X_i^{(b)}|_\rho)_{i\in[m+1],b\in\bits}\approx_{2(m+1)2^{-4k^{.996}}} (U_{.2k^{.996}})_{i\in[m+1],b\in\bits}.\] Therefore, we can upper bound Equation $\ref{eqn:kpartyerror}$ by \[\E_{(Y_i^{(b)})_{i\in[m]},b\in\bits}e\left(\sum_{\delta\in\bits^{m+1}} f(Y_1^{(\delta_1)}),\dots, Y_{m+1}^{(\delta_{m+1})})\right) + 4(m+1)2^{-4k^{.996}}\le R_{m+1}(f) + 4(m+1)2^{-4k^{.996}}\] as desired.
    
\end{proof}

With this, we can show that $f\circ \Ext^{m+1}|_\rho$ uncorrelates against randomized multiparty protocols.

\begin{thm}
\label{thm:extffmvsnof}
Let $g:\bits^n\to \bits$ be a Boolean function, and let $\rho$ be a restriction such that $X_i|_\rho$ has $\ge k^{.998}$ live variables for each $i$, and $g|_\rho$ can be computed by an $(m+1)$-party NOF randomized protocol with with $\le c$ bits and with error $\gamma$. Then $$\corr(f\circ\Ext^{m+1}|_\rho,g|_\rho)\le 2\gamma + 2^{c - \Omega(k^{.996}/2^m)}.$$

\end{thm}

\begin{proof}
Notice that a randomized NOF protocol is simply a distribution over deterministic NOF protocols. Let $\pi$ be some distribution over protocols such that for each $x$, $\Pr_{P\sim \pi}[P(X) \neq g|_\rho(X)] \le \gamma$. We now manipulate the correlation as \begin{align*}
    \corr(f\circ\Ext^{m+1}|_\rho, g|_\rho)  & = |\E_{X}(-1)^{f\circ\Ext^{m+1}|_\rho( X)+ g|_\rho(X)}| \\ &  \le |\E_{X,P}[(-1)^{f\circ\Ext^{m+1}|_\rho(X)+ P(X)}]| + 2\gamma \\ & = 2^c\E_P[R_{m+1}(f\circ\Ext^{m+1}|_\rho)^{1/2^{m+1}}] + 2\gamma \\ 
    & \le 2\gamma + 2^c(R_{m+1}(f) + 4(m+1)\cdot 2^{-.4k^{.996}})^{1/2^{m+1}}
    \\ & \le 2\gamma + 2^c(\corr(f,\Pi_{m+1}^{m+1}) + 2^{-\Omega(k^{.996})})^{1/2^{m+1}} \\ & \le 2\gamma + 2^{c - \Omega(k^{.996}/2^m)}
\end{align*}

\end{proof}

 We now have all the ingredients to finish. Say $\rho$ is good if $\rho$ keeps $\ge k^{.998}$ variables alive in each block $X_i$ \emph{and} $g|_\rho$ is computable by $\{\sym,\thr\}\circ \AND_m$. We have shown for $\rho\sim D$, this doesn't happen only with probability $2^{-\Omega(pk)}$. If $g|_\rho$ has a $\sym$ gate at the top, then Theorem \ref{thm:hg91} says the $\sym\circ\AND_m$ circuit can be computed by a deterministic NOF protocol over $X_1,\dots, X_{m+1}$ using $O(m\log s)$ bits. Plugging this in to Theorem \ref{thm:extffmvsnof} tells us \[\corr(f\circ\Ext^{m+1}|_\rho,g|_\rho)\le 2^{m\log s - \Omega(k^{.996}/2^m)}\le 2^{-\Omega(n^{.995})}.\] If the top gate is a $\thr$, use Theorem \ref{thm:nis93} with $\gamma_{err} = 2^{-n^{.997}}$ to get that the circuit is a randomized NOF protocol over $X_1,\dots, X_{m+1}$ using $O(m^3\log n\log(n/\gamma_{err})) = O(n^{.995})$ bits. Plugging this into Theorem \ref{thm:extffmvsnof} gives us a correlation bound of \[\corr(f\circ\Ext^{m+1}|_\rho,g|_\rho)\le 2^{n^{.995} - \Omega(k^{.996}/2^m)}\le 2^{-\Omega(n^{.996})}.\] In either case we get the same bound, so we can bound \begin{align*}\corr(f\circ\Ext^{m+1},g) & = |\E_{\rho\sim D}\E_X(-1)^{f\circ\Ext^{m+1}|_\rho(X) + g|_\rho(X)}| \\ & \le 2^{-\Omega(pk)} + \E_{\rho\sim D}[|\E_X(-1)^{f\circ\Ext^{m+1}|_\rho(X) + g|_\rho(X)}| | \rho\text{ is good}] \\ & \le 2^{-\Omega(pk)} + 2^{-\Omega(n^{.995})} \\ & = 2^{-\Omega(n^{.995})}. \end{align*} The theorem is proved. 
\end{proof}

\begin{remark}
\label{rem:involved-analysis}
    We note that the original $\rw$ function instantiated with different parameters can also get the same strengthened correlation bound. This requires a more nuanced analysis than present in \cite{st19}, and does not extend to general functions of the form $f\circ \Ext^{m+1}$ as it relies on the specific structure of $\gip$ and $\bigoplus$.
\end{remark}

To recap the argument for a size $s$ circuit, we first use the multi-switching lemma to reduce to a depth-2 circuit of fan-in $\log s$. We then restrict more variables so that the fan-in reduces to $\sqrt{\log s}$. We then apply correlation bounds for $\sqrt{\log s}$-party protocols to get an error of $\exp(-n/2^{\sqrt{\log s}})$. If one trusts that this error is the bottleneck in the argument, one can imagine running through the above argument again with $s=n^{\Theta(1)}$ to get a better error.

\begin{cor}
    \label{cor:poly-refine}
    Let $g(X)$ be a function implementable by a size $s = n^{O(1)}$-size $\{\sym, \thr\}\circ \ac$ circuit, and let $m = 2^{\sqrt{\log n}}$. Define $k:= n/(m+1)$, and let $\Ext:\bits^k\to\bits^{k/2^{O(\sqrt{\log n})}}$ be a $(k/2^{O(\sqrt{\log n})},2^{-k/2^{O(\sqrt{\log n})}})$-extractor constructed from \cite{kz07}. Then $$ \corr(f\circ \Ext^{m+1}, g)\le 2^{-(n/2^{O(\sqrt{\log s})})}.$$ 
\end{cor}

This refinement will be useful for our correlation bounds against branching programs in the next section. As the proof is extremely similar to the above, we defer the sketch to the appendix.

From \Cref{thm:circuit}, we derive the following two theorems as well.

\begin{thm}
\label{thm:circprg}
    There exists an $\eps$-PRG against size-$S$ $\{\sym,\thr\}\circ\ac$ circuits with seed length $s = 2^{O(\sqrt{\log S})} + (\log(1/\eps))^{2.01}$
\end{thm}

\begin{proof}
    First notice that it is trivial to adjust the constants in our argument of Theorem \ref{thm:circuit} so that our correlation is $2^{-\Omega(n^{1-\delta})}$ for arbitrarily small constant $\delta$ Set $r= 2^{O(\sqrt{\log (nS^2)})} + (\log(1/\eps))^{1/(1-\delta)}$ Theorem \ref{thm:circuit} then gives us a function $f:\bits^r\to\bits$ which obtains correlation \[\exp(-\Omega(r^{1-\delta})) = \exp(-\Omega(\log(1/\delta)))\le \eps\] for $\{\sym,\thr\}\circ \ac$ of size \[\le r^{O(\log r)} = 2^{O(\log^2 r)}\le nS^2.\] Since a $\{\sym,\thr\}\circ \ac\circ \junta_{\log n}$-circuit of size $S$ can be trivially expressed by a $\{\sym,\thr\}\circ \ac$ of size $nS^2$, it follows that by Theorem \ref{thm:nw}, we can construct an $\eps n$-PRG against size-$S$ $\{\sym,\thr\}\circ \ac$ circuits with seed length \[O(r^2/\log n) = 2^{O(\sqrt{\log (nS^2)})} + (\log(1/\eps))^{2/(1-\delta)} = 2^{O(\sqrt{\log S})} + (\log(1/\eps))^{2.01}\] for $\delta$ small enough and $S\ge n$. Substituting $\eps\gets\eps/n$ gives the desired result. 
\end{proof}

\begin{lem}
\label{thm:any}
    If $g$ is computable by a $\junta_{n^{.997}}\circ\sym\circ\ac$ circuit where each depth-2 $\sym\circ\ac$ subcircuit is of size $s = n^{O(\log n)}$, then \[\corr(f\circ\Ext^{m+1}, g)\le 2^{-\Omega(n^{.997})}.\] Similarly, if $g$ is computable by a $\junta_u\circ\sym\circ\ac$ circuit where each depth-2 $\thr\circ\ac$ subcircuit is of size $s = n^{O(\log n)}$, then \[\corr(f\circ\Ext^{m+1}, g)\le 2^{-\Omega(n^{.997}/u)} .\]
\end{lem}

\begin{proof}
    We essentially prove a variant of \ref{thm:circuit} where the circuit class is now $\junta_u\circ \{\sym,\thr\}\circ \ac$, an arbitrary $\junta_u$ with size $s = n^{O(\log n)}$ $\{\sym,\thr\}\circ \ac$ circuits hanging from each of the $u$ input wires. $u$ will end up representing the number of $\{\sym,\thr\}$ gates we use. As this circuit is now of size $us$, we see a variant of Lemma \ref{lem:circtodt} will be shown (since this was inherited from \cite{st19}, see Appendix B in the paper for details)

    \begin{lem}
Let $g\in \junta_u\circ \{\sym,\thr\}\circ \ac_d$ with circuit size $s = un^{\tau\log n}$. Then for $p = \frac{1}{48}(48\log s)^{-(d-1)} $ \begin{align*}\Pr_{\rho\gets{\cal R}_p}[g|_\rho &\text{ is not computed by $(.001pk,\junta_u\circ \{\sym_{s^2},\thr_{s^2}\}\circ \AND_{\log s}\})$-tree}] \\ & \le s\cdot 2^{-.001pk/2^d} \\ & = 2^{-\Omega_d(pk)}\end{align*}
\end{lem}
The remainder of the proof is exactly identical as in Theorem \ref{thm:circuit} until the ending regarding multiparty protocols. At this stage, with $2^{-\Omega(n^{.999})}$, $g|_\rho$ has been simplified to a $\junta_u\circ\{\sym,\thr\}\circ\AND_m$ circuit, and has left $\ge k^{.998}$ variables per input block alive. 

\textbf{Case 1 - $\sym$ Gate}: By a trivial extension of Theorem \ref{thm:hg91}, this circuit can be calculated by an $(m+1)$-party protocol using $u\cdot O(m\log s)$ bits (explicitly stated as Fact B.3 in \cite{st19}). Therefore, by Theorem \ref{thm:extffmvsnof}, \[\corr(f\circ\Ext^{m+1}|_\rho,g|_\rho)\le 2^{-\Omega(um\log s-k^{.998}/2^m)}.\] Setting $u = n^{.997}$ gets the desired result. 

\textbf{Case 2 - $\thr$ Gate}: By a trivial extension of Theorem \ref{thm:nis93}, this circuit can be calculated by an $(m+1)$-party protocol with error $\gamma$ using $ u\cdot O(m^3\log n\log(n/\gamma))$ bits (explicitly stated as Theorem 13 in \cite{st19}). Setting $\gamma = 2^{-\Omega(k^{.998}/u)}$, it follows by Theorem \ref{thm:extffmvsnof}, \[\corr(f\circ\Ext^{m+1}|_\rho,g|_\rho)\le 2^{\gamma} + 2^{-\Omega(um^3\log n\log(n/\gamma)-k^{.998}/2^m)} = 2^{-\Omega(k^{.998}/u)},\] yielding the desired result.
\end{proof}

\begin{thm}
    Let $\ac[\g,t,s]$ be the class of size-$s$ $\ac$ where $\le t$ gates are allowed to be of type $\g$. We then have \[\corr(f\circ\Ext^{m+1},\ac[\sym,n^{.996}, n^{O(\log n)}]\le 2^{-\Omega(n^{.996})}\] and \[\corr(f\circ\Ext^{m+1},\ac[\thr,n^{.49}, n^{O(\log n)}]\le 2^{-\Omega(n^{.49})}.\]
\end{thm}

\begin{proof}
    We replicate the proof given in \cite{ls11}. Let $g\in \ac[\g,t,s]$. Let $G_1,\dots, G_t$ be the gates which occur in $g$, sorted such that it respects the topological order (if gate $G_i$ is contained in the subcircuit with top gate $G_j$, then we must have $i < j$). WLOG, assume the top gate is $G_t$. Consider the decision tree $T$ computing $g$ which sequentially queries the value of the circuit $C_i$ defined by the subcircuit with top gate $G_i$ for $i = 1,2,\dots, t$ with the following caveat: if $C_i$ contains some gate $G_j$, we must have already queried that subcircuit, and so replace it with the queried bit. With this caveat, all $C_i$ are $\G\circ\ac$ circuits of size at most $st$. Hence, $T$ is a depth-$t$ decision tree which queries $\g\circ\ac$ circuits at each step. Note for a path $P$ of $T$, notice the predicate $h_P(x)$, which indicates whether an input caused a traversal down path $P$ in $T$, is a $\junta_t\circ G\circ\ac$ function, as it simply checks that all the $t$ circuits queried down that path has the desired output. Using the fact that an input $x$ cannot simultaneously indicate multiple accepting paths, it follows that \[(-1)^{g(x)} = \sum_{\text{accepting }P}(-1)^{h_P(x)} - (N-1)\] where $N$ is the number of accepting paths. Therefore, \[\corr(f\circ\Ext^{m+1},g)\le \sum_{\text{accepting }P}\corr(f\circ\Ext^{m+1},h_P) + N\corr(f\circ\Ext^{m+1},0)\] Noting that $N\le 2^t$, and each $h_P$ along with $0$ can be computed by $\g\circ\ac$ of size $\le st$, it follows from Lemma \ref{thm:any} that for $\g=\sym$, \[\corr(f\circ\Ext^{m+1},\ac[\sym,n^{.99}, n^{O(\log n)}]\le 2\cdot 2^{n^{.996}}\cdot 2^{-\Omega(n^{.997})} = 2^{-\Omega(n^{.997})}\] and if $\g = \thr$,  \[\corr(f\circ\Ext^{m+1},\ac[\thr,n^{.49}, n^{O(\log n)}]\le 2\cdot 2^{n^{.498}}\cdot 2^{-\Omega(n^{.997}/n^{.498})}\le 2^{-\Omega(n^{.499})}.\]
    
\end{proof}

Via a straightforward proof, almost exactly the same as that of Theorem \ref{thm:circprg}, we can use the Nisan-Wigderson framework to turn the above theorem into new PRGs against $\ac$ with a limited number of $\{\sym,\thr\}$ gates.

\begin{thm}
    There is an efficient $\eps$-PRG which fools $\ac[\sym,n^{.999}, S]$ with seed length $2^{O(\sqrt{\log S})} + (\log(1/\eps))^{2.01}$ and an $\eps$-PRG which fools $\ac[\thr,n^{.499},S]$ with seed length $2^{O(\sqrt{\log S})} + (\log(1/\eps))^{4.01}$.
\end{thm}

\section{PRGs against $(d,\poly(n),n)$-2BPs}
In this section, we use fortified hard functions to establish strong correlation bounds against the XOR of juntas, $\junta^{\oplus \poly(n)}_{n,d}$. These are then pushed through the Nisan-Wigderson  ``hardness vs. randomness'' framework to create PRGs which can fool $(d,\poly(n),n)$-2BPs. We first establish the correlation bounds, and then we show that this implies our desired PRG.

\subsection{Correlation Bounds Against $\junta^{\oplus \poly(n)}_{n,d}$} 
This subsection is devoted to proving the following result.

\begin{thm}
\label{thm:juntacorr} Let  $m = d\log n$, let $h$ be the hard function in \Cref{cor:poly-refine} instantiated on $k:= n/m$ bits, and let $\oplus_m:\bits^m\to\bits$ be the parity function on $m$ bits. We then have
    \[\corr(h\circ\oplus_m^k, \junta_{n,d}^{\oplus n^c})\le \exp\left(-\frac{n}{d2^{O(\sqrt{\log n})}}\right) \]
\end{thm}

\begin{proof}
    Consider arbitrary $g\in \junta_{n,d}^{\oplus n^c}$. We will show that there exists a subset $T\subset[n]$ of variables such that upon fixing all variables outside $T$, $g$ simplifies to a sparse polynomial, while at least one input variable in each $\oplus_m$ stays alive. Write $f = \sum_{i=1}^{n^c}\phi_i$, where each $\phi_i$ is a $d$-junta. Let $S_i\subset[n]$ be the indices of the variables that $\phi_i$ depends on. Pick $T\subset_{1/d}[n]$.  For a fixed $i$, we can bound \begin{align*} \Pr_T[|T\cap S_i|\ge \ell] & \le \sum_{\substack{ S\subset S_i \\ |S| = \ell }} \Pr_T[S\subset T] \\ &  = \binom{d}{\ell}\left(\frac{1}{d}\right)^{\ell} \\ & \le  \exp(-\Omega(\ell\log\ell ))\\ & \le 0.1n^{-c}.\end{align*} for $\ell = \Theta(\log n)$. Union bounding over all $i$, it follows that \begin{align}\Pr_{\rho\sim\calR_{1/d}}[\exists i, |T\cap S_i| \ge \ell ] < 0.1.\label{eqn:sparse-simpl}\end{align}

Let $X_1,\dots, X_k$ be the input blocks of size $m$ feeding into $h$. We can easily calculate \begin{align}\Pr_{T}\left[\exists i, X_i\cap T = \emptyset\right]\le k(1-1/d)^m \le k\exp (-m/d) = 1/m = o(1). \label{eqn:entropy-preserve}\end{align} Union bounding \Cref{eqn:sparse-simpl} and \Cref{eqn:entropy-preserve}, it follows that there exists a subset $T\subset[n]$ that simultaneously intersects at most $\ell$ variables alive in each junta $\phi_i$, and intersects at least one variable in each $X_i$. By pruning out elements, we can assume WLOG that there is exactly one variable in each $X_i$. 

Since a function over $b$ bits can be written as an $\F_2$-polynomial with up to $2^b$ terms, it follows for any restriction $\rho$ with $\rho^{-1}(\star) = T$, $\phi_i|_\rho$ is a polynomial with $2^\ell = n^{\Theta(1)}$ terms. Therefore, $f|_\rho$ is a polynomial with $n^{\Theta(1)}$ terms as well, which can be written as a $n^{\Theta(1)}$-sized $\pargate\circ \AND$ circuit. Furthermore, we know that $h\circ \oplus_m^k|_\rho$ is equivalent to $h$ up to negations of the inputs. As $\sym\circ\ac$ is invariant under shifts of the input, we can appeal to \Cref{cor:poly-refine} and observe \begin{align*}\corr(h\circ\oplus_m^k, g) & = |\E_X(-1)^{h\circ\oplus_m^k(X) + g(X)}| \\ & \le  \E_{X_{\conj{T}}}|\E_{X_T}(-1)^{h\circ \oplus_m^k(X_T,X_{\bar{T}}) + g(X_T,X_{\bar{T}})}| \\ & \le \exp\left(-(n/d)/ 2^{O(\sqrt{\log n})}\right)\end{align*}

\end{proof}

\subsection{Constructing and Analyzing the PRG}

With this correlation bound in hand, we can construct good PRGs against the XOR of juntas using the Nisan-Wigderson framework.

\begin{cor}
\label{cor:juntaprg}
    There is an $\eps$-PRG for $\junta^{\oplus n^{\Theta(1)}}_{n, d}$ with seed length $s = 2^{O(\sqrt{\log n})}d^2\log^2(1/\eps))$
\end{cor}

\begin{proof}

    By Theorem \ref{thm:juntacorr} we have an explicit function $f:\bits^r\to\bits$ such that \[\corr(f,\junta^{n^{100}}_{r, d\log n})\le \exp\left(-\frac{r}{2^{O(\sqrt{\log n})}d\log n}\right). \] We can set $r := d2^{\Theta(\sqrt{\log n})}\log(1/\eps)$ so that the above correlation is below $\eps$. Therefore, $f$ is $\eps$-hard for \[\junta^{\oplus n^{100}}_{r, d\log n}\supset \left(\junta^{\oplus n^{100}}_{n, d}\right)\circ \junta_{r,\log n},\] since the composition of an $a$-junta with a $b$-juntas is an $ab$-junta. By Theorem \ref{thm:nw} with $k \gets \log n$ and $r$ as defined earlier, it follows there exists an $\eps$-PRG for $\junta^{\oplus n^{\Theta(1)}}_{n, d}$ with seed length \[O\left(\frac{d^2 2^{\Theta(\sqrt{\log n})}\log^2(1/\eps)}{\log n}\right) = 2^{O(\sqrt{\log n})}\cdot d^2\log^2(1/\eps)\] as desired.
\end{proof}

We now show that fooling the parity of juntas actually allow us to fool arbitrary functions of juntas as long as the function has low Fourier $L_1$ norm.

\begin{thm}
\label{thm:prglift}
    Let $G$ be an $\eps$-PRG for $\junta^{\oplus m}_{n,d}$, and let $f:\bits^m\to \bits$. Then $G$ is an $\eps\cdot L_1(f)$-PRG for $f\circ \junta_{n,d}$.
\end{thm}

\begin{proof}
    Let $g\in f\circ \junta_{n,d}$, implying that $g(x) = f(\phi_1(x),\dots, \phi_m(x))$ for some $\phi_i\in \junta_{n,d}$. By Fourier expanding $f$, it follows that \[g(x) = \sum_{S\subset[m]}\hat{f}(S)(-1)^{\sum_{i\in S}\phi_i(x)}\] Notice that for each $S$, $(-1)^{\sum_{i\in S}\phi_i(x)}\in \junta^{\oplus m}_{n,d}$. Therefore if we let $D$ be the pseudorandom distribution over $\bits^n$ induced by the image of $G$, it follows \begin{align*} |\E_{x\sim D}[g(x)] - \E_{x\sim U_n}[g(x)]| & \le \sum_{S\subset[m]}|\widehat{f}(S)||\E_{x\sim D}[(-1)^{\sum_{i\in S}\phi_i(x)}] - \E_{x\sim U}[(-1)^{\sum_{i\in S}\phi_i(x)}]| \\ & \le \eps\sum_{S\subset[m]}|\widehat{f}(S)| \\ &  = \eps\cdot L_1(f) \end{align*}
\end{proof}

Finally, as an application, we show PRGs against $(d,t,n)$-2BPs, branching programs over $n$ bits with width 2, length $t$, and reads $d$ bits at a time. We will use the fact that width-2 branching programs which read one bit at a time have low Fourier $L_1$ norm (a proof can be found in \cite{hh23}).

\begin{lem}
\label{lem:bpl1}
    If $f$ is a $(1,t,n)$-2BP, then $L_1(f)\le (t+1)/2$.
\end{lem}

We now use the fact that a $(d,t,n)$-2BP can be represented by a normal width-2 branching program acting on juntas to prove that the PRG from Corollary \ref{cor:juntaprg} fools $(d,t,n)$-2BPs.

\begin{thm}
\label{thm:nwbpprg}
    There exists an $\eps$-PRG for $(d, n^c,n)$-2BPs with seed length $s = 2^{O(\sqrt{\log n})}\cdot d^2\log^2(n/\eps)$.
\end{thm}

\begin{proof}
    Given a $(d, n^c,n)$-2BP $B$, we note that at each vertex $v\in[2n^c]$ of $B$, the transition function is some $d$-junta $\phi_v$ which will map the $d$ bits read at that vertex to the next vertex to move to. \ Now consider the $(1,n^c,2n^c)$-2BP $B'$ defined with the same vertex set as $B$, and define the transition function for $v\in[2n^c]$ in $B'$ to read the $v$th bit of the input, and then map to the node in the next layer labeled by that bit. It is easy to see by construction that $B(x) = B'(\phi_1(x),\dots,\phi_{2n^c}(x))$, which is a function in $B'\circ \junta_{n,d}$. By Theorem \ref{thm:prglift}, this can be $\eps$-fooled by an $(\eps/L_1(B'))$-PRG for $\junta^{\oplus 2n^c}_{n,d}$. Using the $L_1$ bound from Lemma \ref{lem:bpl1} and the construction from Corollary \ref{cor:juntaprg}, we see that such a PRG has seed length $2^{O(\sqrt{\log n})}d^2\log^2(1/\eps)$.
\end{proof}

\begin{remark}
    There is an alternative PRG construction using the Ajtai-Wigderson framework \cite{aw85} which gives optimal dependence on $d$, but exponentially worse dependence on $\eps$. This is presented in \Cref{sec:prgaw}
\end{remark}

\section{Correlation Bounds Against Set-Multilinear Polynomials}

Our correlation bound for set-multilinear polynomials follows from an instantiation of the following theorem.
 
\begin{thm}
\label{thm:mintamp}Let $d\le n$ be an integer. Let $\Ext: \bits^{n/d}\times \bits^{2n/d} \to\bits^{k-2\log(1/\eps)}$ be a strong linear seeded $(k,\eps)$-extractor with seed length $2n/d$ created from the Leftover Hash Lemma \cite{ill89}, and let $\chi$ some nontrivial additive character of $\F_{2^{n/d}}$. Define $\extffm_d: \bits^{n+2n/d}\to \bits$ to be \[\extffm_d(X,W) = \chi\left(\prod_{i=1}^d \Ext(X_i, W)\right).\] Let $g:\bits\to\bits^n$ be a function, and let $S_1,\dots, S_{d}\subset [n/d]$ be subsets of size $\ge k$ such that for any restriction $\rho$ created by arbitrarily fixing all bits in $W$ and outside $S_i$ in $X_i$ for each $i$, $g|_\rho$ always becomes set multilinear in $X_1,\dots, X_d$. We then have $$ \corr(\extffm_d,g)\le d\eps + (d-1)\left(\frac{1}{2^k\eps^2}+\eps\right).$$
\end{thm}
\

\begin{proof}
For brevity, we let $f := \extffm_d$ in this proof. We will first split the correlation expectation into first randomizing over all restrictions $\rho$ of the bits in $X$ outside of $S_1,\dots, S_d$, then over the seed $W$, and then over the remaining live variables denoted by the $S_i$, which we denote $X_1|_\rho,\dots, X_d|_\rho$. Now let $W_{\rho}$ be the set of seeds $w$ such that $\Ext(X_{i}|_\rho, w)\approx_\eps U_k$ for all $i$. As $\Ext$ is strong-seeded, it follows by a union bound that $W_{\rho}$ cover all but a $d\eps$ fraction of seeds.  Thus one can write \begin{align}\corr(f,g)  & = |\E_X(-1)^{f'(X)+g(X)}| \nonumber\\ 
& \le \E_{W,\rho}\left|\E_{X}(-1)^{f|_\rho(X,W) + g|_\rho(X,W)}\right| \nonumber\\ 
& \le d\eps + \E_\rho \E_{w\in W_\rho}|\E_X(-1)^{f|_\rho(X,w) + g|_\rho(X,w)}|\label{eqn:goodseed}\end{align} Now fix a partial assignment $\rho$ and seed $w\in W_\rho$. For brevity, let $f(\cdot) := f|_\rho(\cdot, w)$, and similarly for $g'$. By assumption, $g'$ is set-multilinear over $X$ We now apply a similar argument showing up in \cite{bhhkk}. Let $\alpha$ be a map taking linear forms $\sum_{i\in [n/d]}c_{i}X_{d,i}$ in $X_d$ to its vector of coefficients $(c_i)\in\F_2^{n/d}$. Note that by this definition, for any linear form $\ell(X_d)$, $\langle\ell(X_d),X_d\rangle = \ell(X_d)$. Letting $e(x) = (-1)^x$. We then see \begin{align}\left|\E_{X}(-1)^{f'(X) + g'(X)}\right| & = \bigg|\E_{X}e\bigg(f(X_{i}) + \sum_{i\in [d-1]}g_i(X_{-i}) +  g_d(X_d)\bigg)\bigg| \nonumber\\
& \le \E_{X_{[d-1]}}\bigg|\E_{X_d}e\bigg(\langle \alpha(f(X_{i}) + \sum_{i\in [d-1]}g_i(X_{-i})), X_d\rangle +  g_d(X_{-d})\bigg)\bigg|  \nonumber\\ 
& \le \Pr_{X_{[d-1]}}\left[\alpha(f'(X) + \sum_{i\in [d-1]}g_i(X_{-i})) = 0\right]\label{eqn:blah}
 \end{align} 
 where we used the facts that $f'$ is linear in $X_d$ (as $\Ext$ here is a linear seeded extractor), $g_d(X_{-d})$ is independent of $X_d$, and linear forms are perfectly unbiased if their coefficient vector is nonzero. We now repeatedly use the simple inequality that for a linear map $h: \F_2^m\to \F_2^k$ and $a\in \F_2^k$, $\Pr_x[h(x) = a]\le \Pr_x[h(x)=0]$ as follows. 

\begin{align}
   \Pr_{X_{[d-1]}}\left[\alpha(f'(X) + \sum_{i\in [d-1]}g_i(X_{-i})) = 0\right] & = \E_{X_{[d-2]}}\Pr_{X_{d-1}}\left[\alpha(f'(X) +  \sum_{i=1}^{d-2}g_i(X_{-i}))) = \alpha(g_{d-1}(X_{-(d-1)}))\right] \nonumber \\ 
   & \le \Pr_{X_{[d-1]}} \left[\alpha\left(f'(X) +  \sum_{i=1}^{d-2}g_i(X_{-i}))\right) = 0\right] \nonumber\\ 
   & \le \Pr_{X_{[d-1]}} \left[\alpha\left(f'(X) +  \sum_{i=1}^{d-3}g_i(X_{-i}))\right) = 0\right] \nonumber \\ 
   & \le \cdots \nonumber\\ 
   & \le \Pr_{X_{[d-1]}}\left[\alpha(f'(X)) = 0\right] \label{eqn:zeropreimage}
\end{align}

To analyze this probability, we show a quick lemma.

\begin{lem}
\label{lem:coeffzero}
For a linear form $\ell(X_d)$, $\alpha(\ell(X_d)) = 0$ if and only if $\ell(X_d) = 0$ for all $X_d$.
\end{lem}

\begin{proof}
For the forward implication, we simply note that if $\alpha(\ell(X_d)) = 0$, then $\alpha(\ell(X_d))_i = 0$ for every $i$. Consequently, for arbitrary $X_d$, \[\ell(X_d) = \sum_{i=1}^{n/d}\alpha(\ell(X_d))_jX_{d,j}\]
For the reverse implication, say there exists index $i$ such that $\alpha(\ell(X_d))_i \neq 0$. Then notice that if $e_i$ is the unit vector with $1$ in the $i$th index and zero everywhere else, \[\ell(e_i) = \sum_{j=1}^{n/d}\alpha(\ell(X_d))_j(e_i)_j = (\ell(X)d)_i\neq 0.\]
\end{proof}

Therefore, by Lemma \ref{lem:coeffzero}, \[\Pr_{X_{[d-1]}}[\alpha(f'(X)) = 0] = \Pr_{X_{[d-1]}}\left[\forall X_d, \chi\left(\prod_{i=1}^d \Ext(X_i|_\rho,w)\right) = 0\right].\] Clearly if $\prod_{i=1}^{d-1}\Ext(X_i|_\rho,W) = 0$, $f'$ becomes identically zero. When this doesn't happen, the function becomes of the form $\chi(c\cdot \Ext(X_d|_\rho,w))$ for some nonzero $c\in \F_{2^{n/d}}$. We now claim that there must exist some $X_d|_\rho$ such that $\chi(c\cdot \Ext(X_d|_\rho,w))$. Notice that for exactly $2^{n/d-1}$ values of $Y$, $\chi(cY) = 0$. As $w\in W_\rho$, the probability that a random $X_d|_\rho$ has $\Ext(X_d|_\rho, w)$ hit one of these values must be $\ge 1/2 - \eps > 0$, proving the claim. Therefore, in order for $\alpha(f'(X)) = 0$, it is necessary that $\prod_{i=1}^{d-1}\Ext(X_i|_\rho,W) = 0$. Therefore, \begin{align*}\Pr_{X_{[d-1]}} [\alpha(f'(X)) = 0] & \le \Pr_{X_{[d-1]}} \left[\prod_{i=1}^{d-1}\Ext(X_i|_\rho,w) = 0\right] \\ & \le \sum_{i=1}^{d-1}\Pr_{X_i}[\Ext(X_{i}|_\rho,w) = 0] \\ & \le (d-1)\left(\frac{1}{2^{k-2\log(1/\eps)}} + \eps\right)\end{align*}

Stringing the above with inequalities (\ref{eqn:goodseed}), (\ref{eqn:blah}), and (\ref{eqn:zeropreimage}), we find \[\corr(\extffm_d,g)\le d\eps + (d-1)\left(\frac{1}{2^k\eps^2}+\eps\right) \] 

\end{proof}

As a very nice application of this structural theorem, we show that we can achieve exponentially small correlation against $n^{O(1)}$-degree polynomials which are set-multilinear over some partition of the input into up to $n^{1-O(1)}$ parts.

\begin{cor}
\label{cor:corrsetmult}
Let $g$ be a degree $<d$ polynomial which is set-multilinear over an arbitrary partition $(A_1,\dots, A_c)$ of $X$ into $c$ parts. Then $$ \corr(\extffm_d,g) \le  2^{-\Omega(n/cd)}.$$ 
\end{cor}

\begin{proof}
For each $i\in [n/d]$, define $S_i$ to be the largest set among $\{X_i\cap A_1,\dots, X_i\cap A_c\}$ (arbitrarily pick one if there are ties). Notice that the sets $\{X_i\cap A_j\}_{j\in [c]}$ partition $X_i$, and $|X_i| = n/d$. Therefore, we know that each $|S_i|\ge \frac{n/d}c = \frac{n}{cd}$. We now claim that any restriction $\rho$ formed by arbitrarily fixing all the bits in $X_i$ which are outside $S_i$, for each $i$, will make $g|_\rho$ set-multilinear over $(X_1,\dots, X_d)$. Assume for the sake of contradiction there existed some monomial in $g|_\rho(X)$ that contained 2 variables from some $X_i$. Since $S_i\subset X_i$ and $S_j\cap X_i = \emptyset$ for $j\neq i$, both of these variables had to have come from $S_i$. But note that $S_i = X_i\cap A_\ell \subset A_\ell$ for some $\ell$, and we know no monomial has 2 terms from the same $A_i$ by our assumption of $g$. This yields our desired contradiction.\\

Therefore, we can apply Theorem \ref{thm:mintamp} on the sets $(S_i)$ with $k = n/cd$ and $\eps = 2^{-.1n/cd}$ to deduce that $$\corr(f,g)\le d2^{-.1n/cd} + (d-1)(2^{-.8n/cd} + 2^{-.1n/cd}) = 2^{-\Omega(n/cd)}.$$ 
\end{proof}

We can set $c = n^{1-2\delta}$ and $d=n^\delta$ to get $2^{-\Omega(n^\delta)}$ correlation against degree $n^\delta$ polynomials which are set-multilinear over some partition of the input into $n^{1-2\delta}$ blocks, which is interesting as these are exponential correlation bounds over the high degree regime.  From the expressions, we note that there is a tradeoff between the strength of the correlation bounds, the maximum degree polynomials we fool, and the number of parts in our set-multilinear partition we can handle. Unfortunately plugging in $c=n$, which would allow $g$ to be any low-degree polynomial, doesn't yield any nontrivial correlation bound. 

\section{Acknowledgements}
We thank Jesse Goodman and David Zuckerman for helpful discussions. We also thank Jeffrey Champion, Chin Ho Lee, and Geoffrey Mon for comments on an earlier draft of the paper. Finally, we thank anonymous reviewers for comments that greatly improved and simplified the presentation of the paper.

\bibliographystyle{alpha}        
\bibliography{refdb.bib}

\appendix
\section{Addenum to $\{\thr,\sym\}\circ\ac$ Correlation Bounds}

We give a proof sketch of \Cref{cor:poly-refine} here.

\begin{cor}
    Let $g(X)$ be a function implementable by a $n^{O(1)}$-size $\{\sym, \thr\}\circ \ac$ circuit, and let $m = 2^{\sqrt{\log n}}$. Define $k:= n/(m+1)$, and let $\Ext:\bits^k\to\bits^{k/2^{O(\sqrt{n})}}$ be a $(k/2^{O(\sqrt{\log n})},2^{-k/2^{O(\sqrt{\log n})}})$-extractor constructed from \cite{kz07}. Then $$ \corr(f\circ \Ext^{m+1}, g)\le 2^{-n/2^{\sqrt{O(\log n)}}}.$$ 
\end{cor}

\begin{proof}
    We proceed exactly as in \Cref{sec:fatsection} all the way until \Cref{claim:qset}. Here we note that upon setting $s = n^{\Theta(1)}$, we only need to set $q=2^{-\Omega(\sqrt{\log n})}$ to make the desired probability at most some constant. We then continue with the rest of the argument exactly as normal to get the desired correlation bound.
\end{proof}

\section{PRGs for Branching Programs via Ajtai-Wigderson}
\label{sec:prgaw}
Here, we use the Ajtai-Wigderson framework to get a PRG whose is optimal in $d$, but has an exponentially worse dependence on $\eps$ than in \Cref{thm:nwbpprg}. This may serve as proof of concept that we should be able to construct a PRG that is optimal in both parameters.

For strings $x,y\in\bits^n$, define the restriction $x\ostar y\in \{0,1,\star\}^n$ to be \[(x\ostar y)_i = \begin{cases} x_i & y_i = 0 \\ 1 & y_i = 1\end{cases}.\]

We now recall the Ajtai-Wigderson framework.

\begin{thm}[\cite{aw85}, as stated in \cite{hh23}, Theorem 5.3.3]
\label{thm:aw}
    Let $\calF$ and $\calF_{simp}$ be classes of functions $f : \bits^n\to\bits$. Assume that $\calF$ is closed under restrictions.
Let $Z$ be a random variable over $\bits^n$
that can be explicitly sampled using $s$ truly random bits
such that
\[\forall f\in\calF, \Pr[f|_{U\ostar Z} \in \calF_{simp}] \ge 1-\delta\]
where $U\in\bits^n$
is uniform random and independent of $Z$. If $\E[Z_i
] \ge p$ and there is an explicit $\delta$-PRG
for $\calF_{simp}$ with seed length $s'$, then there is a PRG against $\calF$ with seed length $O(p^{-1}\log(n/\delta)(s+s'))$ and error $O(p^{-1}\delta\log(n/\delta))$.
\end{thm}

We first show the following pseudorandom simplification lemma.

\begin{lem}
\label{lem:pseudo-simp}
 For sufficiently large $n$, $\delta\in (0,1)$, and some setting of $\ell = \Theta(\log(n/\delta))$, the following holds. Let $U\sim \bits^n$ be uniformly random and let $Z$ be $(\frac{\delta}{2n^c})$-almost $\ell$-wise close to $\func{Ber}(1/d)^{\otimes n}$. Then it follows for any $f\in \junta_{n,d}^{n^c}$, \[\Pr[f|_{U\ostar Z}\in \junta_{n,\ell)}^{n^c}]\ge 1-\delta.\] Furthermore, $Z$ can be sampled using $O(\log d\log(n/\delta))$ random bits.
\end{lem}

\begin{proof}
Write $f = \sum_{i=1}^{n^c}\phi_i$, where each $\phi_i$ is a $d$-junta. Let $S_i\subset[n]$ be the indices of the variables that $\phi_i$ depends on. We will show that with high probability, only $\log(n/\delta)$ variables in each $S_i$ will stay alive after applying $U\ostar Z$. Letting $\tilde{Z}\sim \func{Ber}(1/d)^{\otimes n}$  \begin{align*} \Pr_Z[\func{wt}(Z_{S_i})\ge \ell] & \le \sum_{\substack{ S\subset S_i \\ |S| = \ell}} \Pr_Z[Z_S = 1^S] \\ & \le  \frac{\delta}{2n^c} + \sum_{\substack{ S\subset S_i \\ |S| = \ell }} \Pr_{\tilde{Z}}[\tilde{Z}_S = 1^S] \\ 
& = \frac{\delta}{n^c} + \binom{d}{\ell}\left(\frac{1}{d}\right)^{\ell} \\ & \le \frac{\delta}{2n^c} + 2^{-\Omega(\ell\log\ell)} \\ & \le \frac{\delta}{n^c}\end{align*} upon setting $\ell = \Theta(\log(n/\delta))$. Union bounding over all $i$ yields that $\Pr[\forall i, \func{wt}(Z_{S_i})< \ell]\ge 1-\delta$. Let $\calE$ be the aforementioned event. Since $\func{wt}(Z_{S_i})$ is exactly the number of variables that stay alive in $\phi_i|_{U\ostar Z}$ (regardless of $U$), $\calE$ implies that each $\phi_i|_{U\ostar Z}$ becomes an $\ell$-junta, which consequently implies $f|_{U\ostar Z}\in \junta_{n,\ell}^{n^c}$. Hence \[\Pr_{U,Z}[f|_{U\ostar Z}\in \junta_{n,\ell}^{n^c}] \ge \Pr_Z[\forall i, \func{wt}(Z_{S_i})< \ell]\ge 1-\delta.\] as desired.

To construct $Z$, we can let $(Z^{(i)})_{1\le i\le \log d}$ be independently sampled $\frac{\delta}{2n^c\log d}$-almost $\ell$-wise uniform strings, and set $Z = \bigwedge_{1\le i\le\log d} Z^{(i)}$. By a simple hybrid argument, it follows $Z$ has the desired distribution. Each $Z^{(i)}$ constructed using $O(\log\ell + \log (2n^c\log d/\delta) + \log\log n) = O(\log(n/\delta))$, and thus $Z$ is constructed using $O(\log d\log(n/\delta))$ bits.
\end{proof}

Next, we show how we can fool $\junta_{n,\Theta(\log(n/\eps))}^{\oplus n^c}$.

\begin{claim}
\label{claim:sparseprg}
    There exists a $\delta$-PRG against $\junta_{n,\Theta(\log(n/\delta))}^{n^c}$ with seed length $2^{O\left(\sqrt{\log(n/\delta)}\right)}$.
\end{claim}

\begin{proof}
    First notice that we can write out a $j$-junta as an $\F_2$-polynomial with $2^j$ terms. Hence, we can write any $f\in \junta_{n,\Theta(\log(n/\delta))}^{n^c}$ as an $\F_2$-polynomial with $n^c\cdot 2^{\Theta(\log(n/\delta))} = \poly(n/\delta))$ terms. Using the sparse polynomial PRG of Servedio and Tan \cite{st22} yields a PRG against this sparse polynomial with seed length $2^{\sqrt{\log(n/\delta)}}$ as desired.
\end{proof}

We now use the Ajtai-Wigderson approach to combine \Cref{lem:pseudo-simp} and \Cref{claim:sparseprg} into a PRG construction.

\begin{thm}
    There exists an explicit $\eps$-PRG against $\junta_{n,d}^{\oplus n^c}$ with seed length $2^{O(\sqrt{\log (n/\eps)})}d$
\end{thm}

\begin{proof}
    Applying \Cref{lem:pseudo-simp} and \Cref{claim:sparseprg} to \Cref{thm:aw}, we get a PRG for $\junta_{n,d}^{\oplus n^c}$ with error $O(\delta d\log(n/\delta))$ and seed length \[O(d\log(n/\delta)(2^{O(\sqrt{\log(n/\delta)})}+\log d\log(n/\delta))) = 2^{O(\sqrt{\log (n/\eps)})}d,\] where we used $d\le n$. Setting $\delta = \Theta\left(\frac{\eps}{2d\log(n/\eps)}\right)$ gives us error $\eps$ and seed length $2^{O(\sqrt{\log (n/\eps)})}d$.
\end{proof}

\end{document}